\documentclass[11pt]{article}
\usepackage{amssymb}
\usepackage{amsfonts}
\usepackage{amsmath}
\usepackage{xcolor}

\numberwithin{equation}{section}

\usepackage{graphicx}

\usepackage[affil-it]{authblk}

\newcommand{\bs}[1]{{\boldsymbol{#1}}}

\def\fddd#1#2{\displaystyle{\frac{\delta #1}{\delta #2}}}

\def\la{\lambda}        
\def\del{\partial} 

\def\bsigma{\bar{\sigma}}
\newcommand{\dsl}[1]{{\displaystyle{#1}}}

\newcommand{\eps}{\epsilon}

\usepackage{graphicx}
\newcommand{\HH}{{{\mathcal{H}}}}
\newtheorem{theorem}{Theorem}[section]

\newtheorem{remark}[theorem]{Remark}

\newcommand{\ol}[1]{{{\overline{#1}}}}
\newcommand{\wit}[1]{{{\widetilde{#1}}}}
\newcommand{\ou}{{\overline{u}}}

\newcommand{\D}{\mathrm{d}}
\newcommand{\drho}{\rho_{{}_\Delta}}
\newcommand\linden{{{\left( h_1
\rho_2+h_2\rho_1 \right)}}}

\newcommand{\bm}[1]{\mbox{\boldmath{$#1$}}}

\renewcommand{\sfdefault}{bch}
\def\barr{\hbox{{\fontfamily{\sfdefault}\selectfont I\hskip -.35ex R}}}
\def\sbarr{\hbox{{\fontfamily{\sfdefault}\selectfont {\scriptsize I}\hskip -.25ex {\scriptsize R}}}}
\def\ssbarr{\hbox{{\fontfamily{\sfdefault}\selectfont {\tiny I}\hskip -.2ex {\tiny R}}}}
\newcommand{\RR}{{\sbarr}}
\newcommand{\colr}[1]{{{\color{black} #1}}}

\begin{document}
\title{Simple two-layer dispersive models\\
in the Hamiltonian reduction formalism}
\author{R. Camassa${}^1$, G. Falqui${}^{2,4,5}$, G. Ortenzi${}^{2,4,6}$, M. Pedroni${}^{3,4}$, T. T. Vu Ho${}^{2,4}$}

\affil{
{\small $^1$University of North Carolina at Chapel Hill, Carolina Center for Interdisciplinary}\\ 
{\small Applied Mathematics, Department of Mathematics, Chapel Hill, NC 27599, USA }\\
{\small camassa@amath.unc.edu}\\
\medskip
{\small  $^2$Department of Mathematics and Applications, 
University of  Milano-Bicocca, \\ Via Roberto Cozzi 55, I-20125 Milano, Italy
}\\
{\small  gregorio.falqui@unimib.it, giovanni.ortenzi@unimib.it, t.vuho@campus.unimib.it}\\
\medskip
{\small $^3$Dipartimento di Ingegneria Gestionale, dell'Informazione e della Produzione, 
\\Universit\`a di Bergamo, Viale Marconi 5, I-24044 Dalmine (BG), 
Italy}\\
{\small marco.pedroni@unibg.it}\\ 
\medskip
{\small  $^4$INFN, Sezione di Milano-Bicocca, Piazza della Scienza 3, 20126 Milano, Italy}\\
\medskip
{\small  $^5$SISSA, via Bonomea 265, 34136 Trieste, Italy} \\
\medskip
{\small $^6$Dipartimento di Matematica ``Giuseppe Peano'' 
Universit\`a di Torino, \\ Via Carlo Alberto 10, I-10123 Torino (TO), 
Italy}\\
{\small  giovanni.ortenzi@unito.it}\\
}

\maketitle

%

\maketitle
\maketitle
\abstract{A Hamiltonian reduction approach is \colr{defined, studied,}  and \colr{finally} used to derive asymptotic models of internal wave propagation in density stratified fluids \colr{in two-dimensional domains}. Beginning with the general Hamiltonian formalism of Benjamin~\cite{Ben86} for an ideal, stably stratified Euler fluid, the corresponding structure is  systematically reduced to the setup of two homogeneous fluids under gravity, separated by an interface and confined between two infinite horizontal plates. A long-wave, small-amplitude asymptotics is then used to obtain a simplified model that encapsulates most of the known properties of the dynamics of such systems, such as bidirectional wave propagation and maximal amplitude travelling waves in the form of fronts. Further reductions, and in particular devising an asymptotic extension of Dirac's theory of Hamiltonian constraints,  lead to the completely integrable evolution equations previously considered in the literature for limiting forms of the dynamics of stratified fluids. \colr{To assess the performance of the asymptotic models, special solutions are studied and compared with those of the parent equations}.
\section{Introduction}
Density stratification in incompressible fluids is an important aspect of theoretical fluid dynamics, and is an inherent component of a wide 
variety of phenomena related to  geophysical application.
Displacement of fluid parcels from their neutral
buoyancy position within a density stratified flow can result in internal wave motion, whose governing equations are not in general amenable to analytical methods for their solutions. Simplified models able to isolate key mechanism of the dynamics that can be studied in detail, even in their one space-dimensional limit,  are therefore valuable and over the years many have been proposed in the literature. A (very) partial list includes~\cite{Chumaetal08,CS93,CGK05,Du16,BB97,Wu98,Wu2000} among many others.

The main focus of our work is 
the study of an ideal stratified fluid from a Hamiltonian viewpoint. The governing equations in the absence of viscosity and diffusivity of the stratifying agent are the Euler equations augmented by density advection, and we consider the simplified case 
consisting of two homogeneous density layers in two spatial dimensions, in the absence of surface tension and confined by two rigid, horizontal, infinite plates.
Hamiltonian aspects of such models with an emphasis on the $2$-layer case have been considered, notably in~\cite{BB97, CS93, CGK05}. 
%
Our approach is an alternative to the one used in~\cite{CGK05} for their two-layer case, and similarly combines asymptotic expansions with the Hamiltonian structure of the original Euler equations. 
However, our starting point is the general density-stratified Hamiltonian of Benjamin~\cite{Ben86},  
and does not make use of the generalization to the two-layer case, as in~\cite{CGK05}, of
Zakharov's Hamiltonian structure~\cite{Zak68} for free surface water waves.  In our approach, once the Hamiltonian reduction~\cite{CFO17} is applied to the two-layer case, we consider different balances between nonlinearity and dispersion, which allows us to retain different asymptotic orders in the ensuing models. In particular, in this paper we focus on a model for interfacial waves propagation between two-homogeneous density fluids for which nonlinearity is stronger than dispersion. This \colr{model, which we shall refer to as the {\em ABC system}, consists of a 
three-paremeter pair of coupled evolution equations} that generalize to bidirectional propagation the well-known KdV-mKdV or Gardner equation for unidirectional motion, and it reduces to it \colr{(together with its Hamiltonian structure)}  through a systematic application of Dirac's Hamiltonian theory of  constraints. In the weakly nonlinear regime, for which a precise nonlinearity and dispersion balance is enforced, the model reduces to the well known integrable cases of Kaup-Boussinesq~\cite{Broer75, Kaup75,Wh2000,Ku85}.   

This paper is organized as follows. Section~\ref{SSEF} is concerned with the details of the derivation of the model equations. Specifically, after a brief review of the fundamental governing equations for ideal, density stratified, incompressible fluids in the section's introduction and in~\S\ref{Sect-1}, we present the elements of Hamiltonian reduction to two-layer flows in~\S\ref{2lay} and~\S\ref{HamredMWR}.   We then proceed to define and apply our asymptotic assumptions in~\S\ref{evHam}-\ref{energ}
to derive the limiting form of the Hamiltonian equations of motion.  Section~\ref{WNLsect}  studies the structure of  
our main model, while the following section, \S\ref{2Red}, considers notable reductions that yield known integrable systems for weakly nonlinear dynamics. Finally, Section~\ref{specsol} considers special solutions that serve to illustrate the models' main features and drawbacks, as well as propose asymptotic equivalences to remedy the latter; lastly, Section~\ref{concper} discusses future directions of investigation and concludes the paper.


%

\section{Density stratified Euler fluids}
\label{SSEF}
 \colr{We consider a perfect, incompressible and variable density fluid confined between two horizontal infinite plates. Thus, the fluid occupies 
 the two-dimensional (2D) domain $(x,z)\in \barr\times(-h_2,h_1)$, with $h_1+h_2\equiv h$ the distance between bottom and upper boundary.
Such a fluid is governed by the incompressible Euler equations for the velocity field $\mathbf{u}=(u,w)$ and non-constant density $\rho(x,z,t)$, in the presence of gravity 
$-g\mathbf{k}$,  
\begin{equation}
\label{EEq}
 \frac{D \rho}{D t}=0, \qquad \nabla \cdot \mathbf{u} =0, \qquad \frac{D (\rho \mathbf{u}) }{D t} + \nabla p + \rho g \mathbf{k}=0 
 \end{equation}
with boundary conditions 
 \begin{equation}
 \mathbf{u}(x=\pm \infty,z,t)=\mathbf{0},\quad  \text{and } w(x,-h_2,t)=w(x,h_1,t)={0},\quad x\in \barr,
 \quad z\in(-h_2,h_1), \quad t\in \barr^+\,, 
 \label{bEEq}
\end{equation}
where $z=-h_2$ and $z=h_1$ are the locations of the bottom and top confining plates, respectively. }
As usual, $D/Dt=\partial/\partial t+\mathbf{u}\cdot\nabla$ is the material derivative.

\subsection{The 2D Benjamin model for heterogeneous fluids in a channel}
\label{Sect-1}
The above system was given a Hamiltonian structure in~\cite{Ben86} with basic, locally measurable variables, i.e., 
the density $\rho$ and  the ``weighted vorticity" $\varsigma$ defined by
\begin{equation}
\label{sigmadef}
\varsigma=\nabla\times (\rho\bs{u})=(\rho w)_x-(\rho u)_z. 
\end{equation}
From~(\ref{EEq}), the equations of motion for these two fields are
\begin{equation}
\label{eqsr}
\begin{array}{l}
\rho_t+u\rho_x+w\rho_z =0\\
\varsigma_t+u\varsigma_x +w\varsigma_z +\rho_x\big(gz-\frac12(u^2+w^2)\big)_z+\frac12\rho_z\big(u^2+w^2\big)_x=0\ .
\end{array}
\end{equation}
These can be written in the form
\begin{equation}
 \label{heq}
{\rho_t}=-\left[\rho,  \dsl{\fddd{\HH}{\varsigma}}\right] \, , \qquad 
\varsigma_t= -\left[\rho,  \dsl{\fddd{\HH}{\rho}}\right]-\left[\varsigma, \dsl{\fddd{\HH}{\varsigma}}\right] \, ,
\end{equation}
where, by definition, $[A, B] = 
A_xB_z-A_zB_x$, and the  functional 
\begin{equation}
\label{ham-ben}
\HH= \dsl{\int_\mathcal{D} \rho\left(\frac12 |\bs{u} |^2+g z\right)\,{\rm d}x\,{\rm d}z}
=\dsl{\int_\mathcal{D} \rho\left(\frac12 |\nabla \Psi|^2+g z\right)\,{\rm d}x\,{\rm d}z}
\end{equation}
is simply given by the sum of the kinetic and potential energy, $\mathcal{D}$ being the fluid domain $\barr\times(-h_2,h_1)$.
The streamfunction $\Psi$ is here used as a placeholder for the map between the weighted vorticity $\varsigma$ and $\bm u$ defined 
by $\varsigma=(\rho \,u)_z-(\rho \,w)_x \equiv -(\rho \, \Psi_z)_z -(\rho \,\Psi_x)_x$. 
As shown in~\cite{Ben86}, equations (\ref{heq}) are a Hamiltonian system with respect to a 
linear Hamiltonian structure, that is, 
they can be written as
\[
 \rho_t=\{\rho, \HH\}
,\qquad \varsigma_t=\{\varsigma, \HH\}
\]
for the Poisson brackets defined by the Hamiltonian operator
\begin{equation}\label{B-pb}
J_B=-
\left(\begin{array}{cc}
       0 & \rho_x \partial_z -\rho_z \partial_x \\ 
       \rho_x \partial_z -\rho_z \partial_x & \varsigma_x \partial_z -\varsigma_z \partial_x
      \end{array}
\right).
\end{equation}

\subsection{Two-layer case}
\label{2lay}
A simplification of system (\ref{EEq})  which retains the essential properties of stratification can be obtained by considering a system of two fluids of homogeneous densities $\rho_2>\rho_1$ in the channel $\barr\times (-h_2, h_1)$. 
\begin{figure}[t]
\centering
\includegraphics[width=12cm]{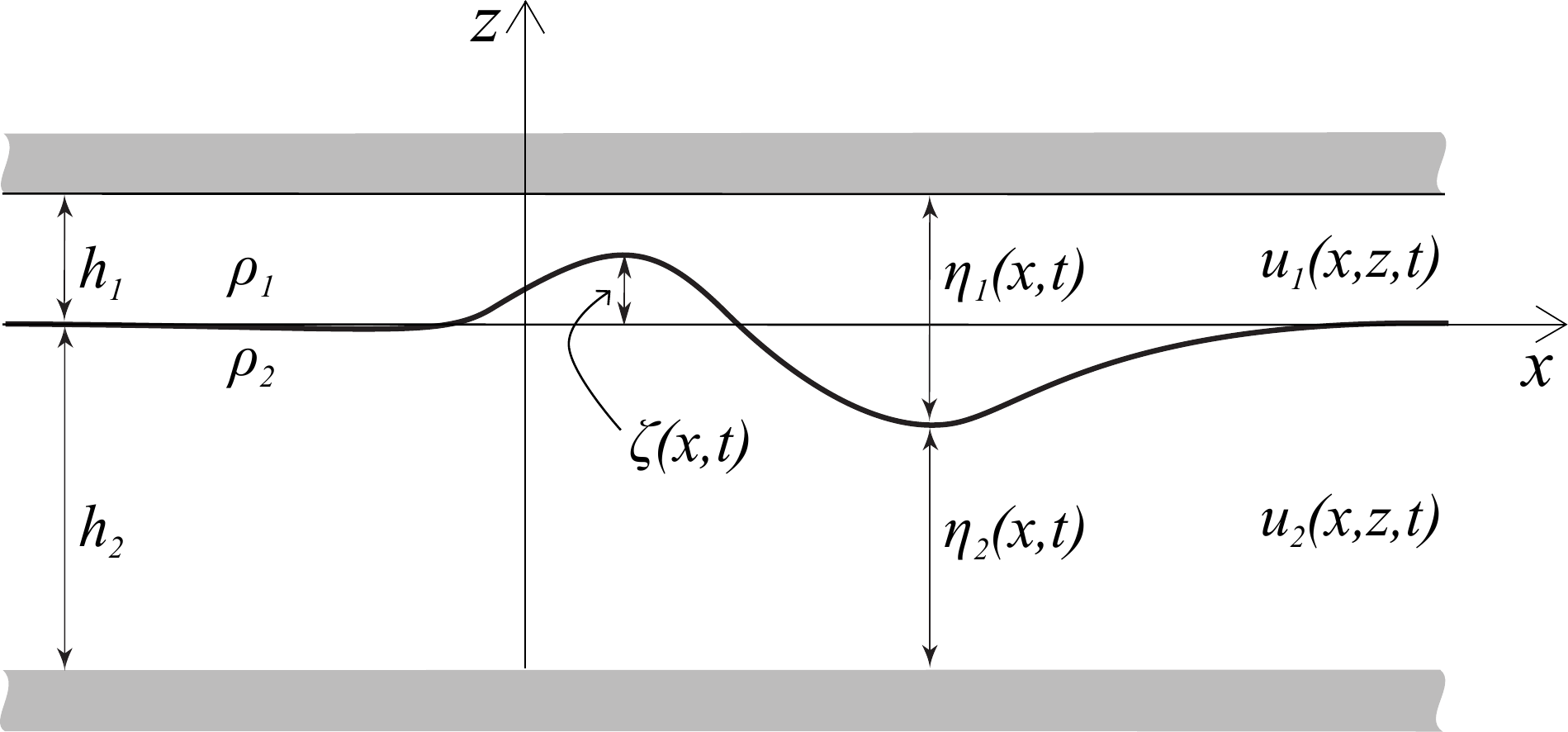}
\caption{Schematics of the two-layered configuration. The quantity $\eta_2$ (resp., $\eta_1$) is the total thickness of the lower, heavier (resp., upper, lighter) fluid. The interface  $\zeta$ is measured from the quiescent state $z=0$. }
\label{addfig}
\end{figure}

The interface 
between the two homogeneous fluids is described by a smooth function $\zeta=\zeta(x,t)$ \colr{(see Figure~\ref{addfig}).}
In this case  the density and velocity fields can be described as 
\begin{equation}\label{no1}
\begin{split}
&\rho(x,z,t)=\rho_2+(\rho_1-\rho_2)\theta(z-\zeta(x,t))\\
&u(x,z,t)=u_2(x,z,t)+(u_1(x,z,t)-u_2(x,z,t))\theta(z-\zeta(x,t))\\
&w(x,z,t)=w_2(x,z,t)+(w_1(x,z,t)-w_2(x,z,t))\theta(z-\zeta(x,t))\, ,\\
\end{split}
\end{equation}
where $\theta$ is the Heaviside function.

A nowadays standard way to reduce the dimensionality of the model is to introduce the layer-averaged velocities  
as set forth by Wu~\cite{Wu81}, since in the case of fluids stratified by gravity the vertical direction plays 
a  distinguished role. Let us denote by
\begin{equation}
\ou_1(x,t)=\frac{1}{\eta_1(x,t)}\int_{\zeta}^{h_1} u_1(x,z,t)\,\D z, \quad \ou_2(x,t)=\frac1{\eta_2(x,t)}\int_{-h_2}^{\zeta} u_2(x,z,t)\, \D z\, ,
\label{layav}
\end{equation}
the layer-averaged velocities, where $\eta_1=h_1-\zeta,\, \eta_2=h_2+\zeta$ are the thicknessess of the layers. Letting  $P(x,t)$ denote the interfacial pressure, the non-homogeneous incompressible Euler equations (\ref{EEq}) 
result in the (non-closed) system 
\begin{equation}
\label{2layer}
 \begin{split}
 & {\eta_i}_t+(\eta_i\ou_i)_x=0, \qquad i=1,2, \\ &
 {\ou_1}_t+{\ou_1}{\ou_1}_x -g {\eta_1}_x + \frac{P_x}{\rho_1} + D_1 =0, \\
 &  {\ou_2}_t+{\ou_2}{\ou_2}_x +g {\eta_2}_x + \frac{P_x}{\rho_2} + D_2 =0.
 \end{split}
\end{equation}
The terms $D_1$, $D_2$ at the right hand side of system~(\ref{2layer}) are 
\begin{equation} 
D_i= \frac{1}{3 \eta_i} \partial_x [\eta_i^3 ({\ou_i}_{xt} +{\ou_i}{\ou_i}_{xx}-({\ou_i}_x)^2)] + \dots, \qquad{i=1,2}\, , 
\label{disperterm}
\end{equation}
where dots represent terms with nonlocal dependence on the averaged velocities. These terms collect the non-hydrostatic correction to the pressure field, and make the evolution of system~(\ref{2layer}) dispersive.
When an asymptotic expansion  based on the long-wave assumption 
$\epsilon \equiv \max[\eta_i/L] \ll 1,\, i=1,2$, is carried out (where $L$ is a typical wavelength), expressions~(\ref{disperterm}) explicitly define the leading order dispersive terms in the small parameter $\epsilon$; truncating at this order makes equations~(\ref{2layer}) local in the layer averaged velocities, resulting in the strongly nonlinear system studied in, e.g.,~\cite{CC99}.

It is important to notice that the first two equations in (\ref{2layer}), which have the meaning of mass conservation laws, 
\begin{equation}
\label{exmce0}
\eta_{j\,t}+\partial_x(\eta_j\,\ou_j)=0\, ,  
\end{equation}
are actually the counterpart of the kinematic boundary conditions at the interface.
\colr{Denoting (here and in what follows)  interface velocities by $\wit{u}_j(x,t)=u_j(x,\zeta(x,t), t)$ and $\wit{w}_j(x,t)=w_j(x,\zeta(x,t),t)$, this can be} seen from the chain of equalities (with $j=2)$,
\begin{equation}
\begin{split}
&\eta_{2\,t} +\partial_x\int_{-h_2}^\zeta u_2(x,z,t)\, \D z=\zeta_t+\zeta_x\wit{u}_2+ \int_{-h_2}^\zeta u_{2\,x} (x,z,t)\, \D z=\\
&(\text{using } u_{2\,x}+
w_{2\,z}=0) =\zeta_t+\zeta_x\wit{u}_2-
\int_{-h_2}^\zeta w_{2\,z} (x,z,t)\, \D z\\
&=\zeta_t+\zeta_x\wit{u}_2-
w_2\Big\vert_{-h_2}^\zeta=(\text{by the bottom no flux condition})\\& =
\zeta_t+\zeta_x\wit{u}_2-
\wit{w}_2=0\,,
\end{split} 
\end{equation}
and similarly for the upper fluid when $j=1$.
Equations (\ref{2layer}) come equipped with two constraints. Namely, we have the obvious geometrical constraint
$\eta_1+\eta_2=h\,$
and its consequence obtained by summing the equations in the first line of (\ref{2layer}), 
\begin{equation}
\label{ddynacon}
 (\eta_1 \ou_1 + \eta_2 \ou_2)_x=0 \, .
 \end{equation}
We remark that under suitable far-field boundary conditions (such as vanishing \colr{or identical} velocities for  $x\to\pm \infty$) this relation translates into the {\em dynamical} constraint
 \begin{equation}
 \label{dynacon}
\eta_1\ou_1+\eta_2\ou_2=0\, .
\end{equation}

\subsection{The Hamiltonian reduction process}
\label{HamredMWR}

We now discuss how a simple averaging process can be given a Hamiltonian structure, 
well suited to the discussion of the constrained equations in which  
our set of reduced coordinates naturally appears. We follow the setting already introduced in \cite{CFO17} and provide a full geometric description of the reduction process.
We begin with definitions~(\ref{no1}), where we suppress time dependence for ease of notation in what  follows.
The two momentum components are 
\begin{equation}\begin{split} 
&\rho u= \rho_2 u_2(x,z)+(\rho_1 u_1(x,z)-\rho_2 u_2(x,z))\theta(z-\zeta(x))\, , 
\\
&\rho w= \rho_2 w_2(x,z)+(\rho_1 w_1(x,z)-\rho_2 w_2(x,z))\theta(z-\zeta(x)) \, , 
\end{split}
\end{equation}
so that 
the weighted vorticity (\ref{sigmadef})
is
\begin{equation}\label{no-s}
\begin{split}
\varsigma=&\rho_2( w_{2\,x}-u_{2\,z})+\big(\rho_1 (w_{1\,x}-u_{1\,z})-\rho_2( w_{2\,x}-u_{2\,z})\theta(z-\zeta(x))
\\&-\big(\rho_1 u_1(x,z)-\rho_2 u_2(x,z)+\zeta_x(\rho_1 w_1(x,z)-\rho_2 w_2(x,z))\big)\delta(z-\zeta(x))\, ,
\end{split}
\end{equation}
where $\delta(\cdot)$ is the Dirac delta function.

We assume that  the motion in each layer is  irrotational, 
so that we are left with a ``momentum vortex line" along the interface, that is, 
\begin{equation}\label{no=s2}
\varsigma=\big(\rho_2 u_2(x,z)-\rho_1 u_1(x,z)+\zeta_x(\rho_2 w_2(x,z)-\rho_1 w_1(x,z))\big)\delta(z-\zeta(x)).
\end{equation}
We define a  projection map 2D $\to$ 1D 
as
\begin{equation}\label{mappazza}
 \zeta(x)=\dsl{\frac{1}\drho}
 \int_{-h_2}^{h_1} (\rho(x,z)-\rho_1)\, \D z \, -h_2, \qquad
{\sigma}(x)=
\int_{-h_2}^{h_1} \varsigma(x,z)\, \D z \, ,
\end{equation}
where $\drho=\rho_2-\rho_1$.
When applied to $2$-layer configurations,  the first of these relations is easily obtained from the first of equations (\ref{no1}). 
Moreover, in the $2$-layer bulk irrotational case, 
\begin{equation}
{\sigma}(x)
=
\rho_2 u_2(x,\zeta(x))-\rho_1 u_1(x,\zeta(x))+\zeta_x(x)(\rho_2 w_2(x,\zeta(x))-\rho_1 w_1(x,\zeta(x))) \, ,
\label{intersi}
\end{equation}
i.e., the averaged weighted vorticity
$
\sigma$ is 
the 
tangential momentum shear at the interface. 

The geometry thus far outlined fits the Hamiltonian reduction scheme devised in~\cite{MR86}. Indeed, such a scheme considers a  manifold $\cal P$ endowed with a Poisson tensor, such as $J_B$,  a submanifold $\mathcal{M}\subset {\cal P}$, a distribution $D$ contained in the tangent bundle to $\cal P$ restricted to $\mathcal{M}$, $T {\cal P}{\vert_\mathcal{M}}$, and state that a Poisson reduction to $\mathcal{M}/\Phi$, with $\Phi$ denoting the intersection $T\mathcal{M}\cap D$, is possible when (some geometrical assumptions on the regularity of $D$ and on its action on $\mathcal{M}$ being granted)
\begin{enumerate}
\item $J_B$ is invariant under $D$.
\item 
At each point of $\mathcal{M}$ it holds	
\begin{equation}\label{MRcond}
J_B(D^0)\subset T\mathcal{M}+ D\, ,
\end{equation}
 \end{enumerate}
$D^0\subset T^*{\cal P}{\vert_\mathcal{M}}$ being the annihilator of $D$ in the cotangent bundle to $\cal P$ restricted to $\mathcal{M}$.

In particular (see the example in~\cite{MR86}), in our case we identify the following geometric objects: 
\begin{enumerate}
\item $\cal P$ is the configuration space  $M^{(2)}$ of 
the 2D fields, parametrized by $(\rho(x,z), \varsigma(x,z))$, and 
$J_B$ is the Benjamin Poisson tensor~(\ref{B-pb})
\begin{equation}\label{B-pbr}
J_B=-
\left(\begin{array}{cc}
       0 & \rho_x \partial_z -\rho_z \partial_x \\ 
       \rho_x \partial_z -\rho_z \partial_x & \varsigma_x \partial_z -\varsigma_z \partial_x
      \end{array}
\right).
\end{equation}
\item  $\mathcal{M}$ is given by the 2-layer configuration space
 \begin{equation}\label{intermfld}
\{\rho(x,z)=\rho_2-
\drho\theta(z-\zeta(x)), \, 
\varsigma(x,z)=
\sigma(x)\delta(z-\zeta(x))\, \}.
\end{equation}
\item $D$ is the image under $J_B$ of the annihilator $T\mathcal{M}^0$ of the tangent space to $\mathcal{M}$ in $TM^{(2)}\vert_{\mathcal{M}}$.
\end{enumerate}
To show how our model fits the Marsden--Ratiu scheme we first notice that the $T\mathcal{M}$ can be described as the space of pairs of generalised functions of the form
\begin{equation}
\label{TI}
\{\dot{\rho}=\drho\dot{\zeta}\delta(z-\zeta),\, \dot\varsigma=\dot{\sigma}\delta(z-\zeta)-\sigma\dot{\zeta} \delta'(z-\zeta)\}\, ,
\end{equation}
$\delta'
$ being the derivative of the Dirac's-$\delta$ function. 
Notice the link between the $\delta(z-\zeta)$-coefficient of $\dot{\rho}$ and the $\delta'(z-\zeta)$-coefficient of $\dot{\varsigma}$.
The annihilator $T\mathcal{M}^0$ is readily computed as pairs of smooth functions $(\phi(x,z),\psi(x,z))$ satisfying 
\begin{equation}
\label{annih-eq}
\psi(x, \zeta)=0,\quad \drho\, \phi(x, \zeta)+\sigma\psi_z(x,\zeta)=0\, .
\end{equation}
Since on $\mathcal{M}$ we have
\begin{equation}
\label{rhozsigmaz}
\begin{array}{ll}
\rho_x=\drho\zeta_x\delta(z-\zeta)\,,&\rho_z=-\drho\delta(z-\zeta)\,,\\
\varsigma_x=\sigma_x\delta(z-\zeta)-\sigma\zeta_x\delta'(z-\zeta)\,,&\varsigma_z=\sigma\delta'(z-\zeta)\, , 
\end{array}
\end{equation}
the restriction of the Poisson tensor $J_B$ on $\mathcal{M}$ acquires the form
\begin{equation}
\label{B-pbI}
J_B\vert_{\mathcal{M}}=-\left(
\begin{array}{cc}\medskip
0 &\delta(z-\zeta) \drho(\zeta_x\partial_z+\partial_x)\\
\delta(z-\zeta) \drho(\zeta_x\partial_z+\partial_x)&\delta(z-\zeta)\,\sigma_x \partial_z-\delta'(z-\zeta)\sigma\,(\zeta_x\partial_z+\partial_x)
\end{array}
\right)\, .
\end{equation}
Hence the image of $ J_B\vert_{\mathcal{M}}$ is the space of vectors 
\begin{equation}
\label{ImJB}
\left(
\begin{array}{l}\medskip
\dot{\rho}\\
\dot\varsigma
\end{array}
\right)=-\left(
\begin{array}{c} \medskip\drho \left(\zeta_x\psi_z+\psi_x\right)\delta(z-\zeta)\, \\
\  \left(\drho\,(\zeta_x\phi_z+\phi_x)+\sigma_x\psi_z\right)\delta(z-\zeta)\,- \sigma\, (\zeta_x\psi_z+\psi_x)\delta'(z-\zeta)\,
\end{array} 
\right)\,.
\end{equation}
This expression can be used to show that $D=J_B(T\mathcal{M}^0)$ reduces to the null vector. 
In fact, let us consider  (\ref{ImJB})  with $(\phi, \psi)$ in $T\mathcal{M}^0$. Thanks to  the $\delta$-function factor, the first component of (\ref{ImJB}) can be written as $\dot{\rho}=-\drho\left(\zeta_x\psi_z(x,\zeta)+\psi_x(x,\zeta)\right)\delta(z-\zeta)$, and the coefficient of the $\delta$  vanishes being the total $x$-derivative of the first of (\ref{annih-eq}).
By using the generalised function  identity $f(y)\delta'(y)= f(0)\delta'(y)-f'(0)\delta(y)$ the second component of (\ref{ImJB}) can be written as
\begin{eqnarray}
\nonumber
&&\hspace{-1cm}
\dot{\varsigma} = \sigma\, (\zeta_x\psi_z(x, \zeta)+\psi_x(x, \zeta))\delta'(z-\zeta)
\\
&&
-\Big(\drho\,(\zeta_x\phi_z(x,\zeta)+\phi_x(x, \zeta))+\sigma_x\psi_z(x, \zeta)+\sigma\zeta_x\psi_{zz}(x,\zeta)+\sigma\psi_{x z}(x,\zeta)\Big)\delta(z-\zeta)\, ,
\label{ImJB2}
\end{eqnarray}
which vanishes as well thanks to (\ref{annih-eq}).
The vanishing of $D$ confirms that the reduced manifold is isomorphic to the submanifold $\mathcal{M}$, which guarantees the invariance of $J_B$. 
As for the characteristic condition for reduction, i.e., equation~(\ref{MRcond}), this follows explicitly from~(\ref{ImJB}) which displays  how the image  $ J_B\vert_{\mathcal{M}}$ is contained in $T\mathcal{M}$ as determined in equation~(\ref{TI}).

We can now compute the expression of the reduced Poisson tensor as follows. We consider the pull-back to  $M^{(2)}$ of  a generic 1-form $(\mu_\zeta(x), \mu_{{\sigma}}(x))$ on the manifold $M^{(1)}$, parametrized by $(\zeta(x),\sigma(x))$, under the map (\ref{mappazza}), given by 
 \begin{equation}
 \label{lift}
 \left(
 \frac1{\drho}
 \mu_\zeta(x),\> 
 \mu_{
 {\sigma}}(x)\right) \, .
 \end{equation} 
 Applying the Poisson tensor (\ref{B-pb}) evaluated on $\mathcal{M}$ to this covector, we obtain 
\begin{equation}\label{PS2}
\left( 
 \begin{array}{c}\bigskip
 \dot\rho(x,z)\\ 
\dot\varsigma(x,z)
 \end{array}\right)
 =-\left( 
 \begin{array}{c}\medskip
 \drho\delta(z-\zeta(x))\left( 
 \mu_{
 {\sigma}}(x)\right)_x
 \\
\delta(z-\zeta(x))\left(
\mu_\zeta(x)\right)_x-
{\sigma}(x)
\delta'(z-\zeta(x))
\left(\mu_{
{\sigma}}(x)\right)_x
 \end{array}\right)\, .
 \end{equation} 
 Pushing this vector to $M^{(1)}$ via the tangent map to (\ref{mappazza}),
 \[
  \dot{\zeta}=\frac1{\drho}\int_{-h_2}^{h_1} \dot{\rho}(x,z)\,\D z, \quad \dot{
  \sigma}=
  \int_{-h_2}^{h_1}\dot{\varsigma}(x,z)\, \D z,
 \]
 yields the vector
\[
(\dot{\zeta},\dot{
{\sigma}})=\left(-
\partial_x\,\mu_{
{\sigma}}, -
\partial_x\,\mu_{\zeta}\right),
\]
owing to the fact that $\int_{-h_2}^{h_1} \delta'(z-\zeta(x))\,\D z =0$ (we work under the assumption that	the fluid interface never touches the boundary, i.e., the strict inequalities $-h_2<\zeta<h_1$ hold).
 
Hence, 
the expression of the reduction of the Benjamin Poisson tensor $J_B$ on the manifold $M^{(1)}$   is given in the coordinates $(\zeta(x), \sigma(x))$ by  the constant  tensor
\begin{equation}\label{Pred}
J_{\rm red}
=-
\left(\begin{array}{cc}
0&\partial_x\\
\partial_x&0\end{array}
\right)\,  .
 \end{equation}
  \colr{This structure coincides with the one introduced in~\cite{BB97}  by a direct inspection of the Hamiltonian formulation of two-layer models. We stress that within our setting the above Poisson tensor is obtained by the process of Hamiltonian reduction from the Lie-Poisson structure of the general heterogeneous incompressible Euler $2D$ fluids of \cite{Ben86}. Moreover, by means of our choice of reducing map (\ref{mappazza}), we  directly obtain a set of coordinates $(\zeta,\sigma)$ that can be called {\em Darboux} coordinates, since they are the analog of the 
coordinates $(u,v)$  for the non-linear wave equation in $1+1$ dimensions 
$ u_{tt}=F''(u) u_{xx}$
derived from the Hamiltonian functional ${\mathcal H}=\frac12\int_{\ssbarr} (u_t^2+F(u)) \, \D x $ by means of the Poisson structure (\ref{Pred}). } 
 

\subsection{The evolution variables and the Hamiltonian}
\label{evHam}
The basic feature of the Hamitonian reduction process is that with this approach 
 the natural dependent variables  are the displacement  from the equilibrium position $\zeta$ 
 and the tangential interface momentum shear
\begin{equation}\label{sigmadef2} \begin{split} 
\colr{{\sigma(x)}}&=\colr{\rho_2 u_2(x,\zeta(x))-\rho_1 u_1(x,\zeta(x))+\zeta_x(x)(\rho_2 w_2(x,\zeta(x))-\rho_1 w_1(x,\zeta(x)))}\\ 
& \equiv\rho_2\wit{u}_2(x)-\rho_1\wit{u}_1(x)+\zeta_x(x)(\rho_2\wit{w}_2(x)-\rho_1\wit{w}_1(x))\, \end{split}
\end{equation}
(\colr{we recall and use hereafter  that} 
a tilde over a quantity 
\colr{stands} for its evaluation at the interface, e.g., 
%
$
\wit{u}_1(x,t)=u_1(x, \zeta(x,t), t)$ etc.).

In this respect, the 
approach we  pursue here differs from the Green-Naghdi setting of,  e.g., \cite{CC99},  which considers {\em layer averaged} velocities (following the seminal paper \cite{Wu81}). Specifically, here we shall  use and adapt to our case the setting discussed in 
\cite{Wu2000} (see also \cite{Zak68, BB97}) where the equations for internal wave motion are written using two sets of coordinates:
\begin{description}
\item[ i)] the {\em boundary velocity basis}, in which $\zeta$ is complemented by $u_{0\,1}(x,t)=u_1(x,h_1,t)$ in the upper layer and by 
$u_{0\,2}(x,t)=u_2(x,h_2,t)$ in the lower layer.
\item[ii)] the {\em interface velocity basis}, where we use the variables entering the Hamiltonian reduction process, that is 
$\wit{u}_j(x,t)=u_j(x, \zeta(x,t), t)$.  
%
%
%
\end{description}

While for some aspects of the theory it is advantageous to use layer-mean velocities (see \cite{Wu2000,CaCh16, CC99}), as mentioned above these are not the ones most naturally suggested by our Hamiltonian reduction procedure, and therefore we choose to express energy, the mass conservation as well as the ensuing dynamical constraint in terms of interface variables.

Following \cite{Wu2000, Wh2000}, we use the assumed bulk irrotationality of the fluid flow to introduce the bulk velocity potentials $\varphi_j(x,z)$, which we Taylor expand with respect to the vertical variable $z$.
By the vanishing of the vertical velocity at the physical boundaries $z=h_1$, and $z=-h_2$ we obtain 
the Taylor expansions
\begin{equation}
\label{pot-j}
\varphi_j(x,z)=\sum_{n=0}^\infty \frac{(-1)^n}{(2n)!} 
H_j(z)
^{2n} \partial_x^{2n}\varphi_{0\, j}(x)\, 
\end{equation}
where 
\begin{equation}
\label{Hdef}
H_1(z)=z-h_1, \quad H_2(z)=z+h_2\, ,
\end{equation}
and $\varphi_{0\,1}(x)=\varphi_1(x,h_1)$, $\varphi_{0\, 2}=\varphi_2(x,-h_2)$ are the values of the potential at the rigid lids.
 
The horizontal velocities are then given by
\begin{equation}\label{u-exp1}
u_j=\partial_x\varphi_j(x,z)=\sum_{j=0}^\infty \frac{(-1)^n}{(2n)!} 
H_j(z)
^{2n} \partial_x^{2n}\partial_x\varphi_{0\, j}(x)=\sum_{j=0}^\infty \frac{(-1)^n}{(2n)!} 
H_j(z)
^{2n} \partial_x^{2n} u_{0\, j}(x)\, , 
 \end{equation} 
$u_{0\, j}(x)$ being the horizontal velocities at $z=h_1$ (for $j=1$) and at $z=-h_2$ (for $j=2$).

Likewise, the vertical velocities are given by 

\begin{equation}\label{w-exp1} 
w_j(x,z)=\partial_z\varphi_j(x,z)=\sum_{n=0}^\infty\frac{(-1)^{n+1}}{(2n+1)!}H_j(z)
^{2n+1} \partial_x^{2n+1}u_{0\, j}(x)
\end{equation}
Notice that the boundary conditions $w_1(x,h_1)=w_2(x,-h_2)=0$ are satisfied.

%
%
Since \begin{equation}
\label{ex-betaj}
H_1(\zeta)=-\eta_1, \quad H_2(\zeta)=\eta_2,\quad\text{i.e., } H_j(\zeta)=(-1)^j \eta_j,\, j=1,2\, , 
\end{equation}
where $\eta_1(x)=h_1-\zeta(x)$ (resp.\ $\eta_2(x)
=h_2+\zeta(x)$) is the thickness of the upper (resp.\ lower) layer,  the interface velocities can be directly obtained by formulas (\ref{u-exp1}) and (\ref{w-exp1}) as
\begin{equation}
\label{intf-speed}
\wit{u}_j=\sum_{j=0}^\infty \frac{(-1)^n}{(2n)!} 
\eta_j^{2n} \partial_x^{2n} u_{0\, j}(x)\, , \qquad \wit{w}_j=(-1)^{j-1}\sum_{n=0}^\infty\frac{(-1)^{n}}{(2n+1)!}\eta_j^{2n+1} \partial_x^{2n+1}u_{0\, j}(x)\, .
\end{equation}
 
For later use, we express (from the same formulas) the layer-mean horizontal velocities in terms of the fluid thicknesses and the (respective) boundary velocities as
\begin{equation}
\label{mean-vel}
\begin{split}
\ou_1(x)&\equiv \frac{1}{\eta_1}\int_\zeta^{h_1} u_1(x,z)\,\D z=\sum_{n=0}^\infty\frac{(-1)^n}{(2n+1)!}\eta_1(x)^{2n}\partial_x^{2n}u_{0\, 1}(x)\, \\ 
\ou_2(x)&\equiv \frac{1}{\eta_2}\int_{-h_2}^\zeta  u_2(x,z)\,\D z=\sum_{n=0}^\infty\frac{(-1)^n}{(2n+1)!}\eta_2(x)^{2n}\partial_x^{2n}u_{0 \,2}(x)\, .
\end{split}
\end{equation}

\subsection{ Rescaling the spatial independent variables: the $\epsilon$--expansion and the mass conservation laws}
\label{rescvar}
To make the formal Taylor series (\ref{intf-speed},\ref{mean-vel}) effective in the construction of asymptotic models for interfacial wave motion we have to rescale variables (see, e.g., \cite{Wh2000, Wu2000}). In particular, we set
\begin{equation}
\label{xz-scale}
x=L\, x^*, \quad z=h\, z^*\, , 
\end{equation}
where $L$ is a typical horizontal scale (say, a typical wavelength) and $h$ is the total height of the vertical channel. As usual, we assume that the ratio $\epsilon=\dsl{{h}/{L}}$ be the small dispersion parameter of the theory.
 Indeed, by using these scalings, we can turn the Taylor series (\ref{u-exp1},\ref{w-exp1}) as well as (\ref{intf-speed},\ref{mean-vel})
into asymptotic series in the small parameter $\eps$. 

For the sake of simplicity, hereafter we shall drop asterisks from the formulas. We remark that,  unless otherwise explicitly stated, horizontal lengths are scaled by $L$ and vertical lengths by $h$.  
Henceforth, we will abuse notation a little and use the order symbol $O(\cdot)$ to denote the magnitude of bounded dimensional 
quantities whenever this can be done without generating confusion.

For the velocity fields we have 
\begin{equation}
\label{uw-eps}
\begin{split}
u_j(x,z)&=\sum_{j=0}^\infty \frac{(-1)^n}{(2n)!} \eps^{2n}
{H_j}(z)
^{2n} \partial_{x}^{2n} u_{0\, j}(x)\,, \\
w_j(x,z)&=(-1)^{j-1}\eps\, \sum_{n=0}^\infty\frac{(-1)^{n}}{(2n+1)!}
\eps^{2n}
H_j(z)^{2n+1} \partial_{x}^{2n+1}u_{0\, j}(x)\, .
\end{split}
\end{equation}
It is worth taking into account here and below the expected (Lagrangian) scaling of vertical vs.\ horizontal velocities $w_j/u_j=O(\eps)$.
Similarly we have
\begin{equation}
\begin{split}\label{intf-vel-eps}
\wit{u}_j&=\sum_{j=0}^\infty \frac{(-1)^n}{(2n)!} \eps^{2n}
{\eta_j}^{2n} \partial_{x}^{2n} u_{0\, j} =u_{0\, j}-\frac{\eps^2}{2}{\eta_j^2}\, u_{0\,j\, xx}+O(\eps^4)\\
\wit{w}_j&=(-1)^{j-1}\eps \, \sum_{n=0}^\infty\frac{(-1)^{n}}{(2n+1)!}\eps^{2n} {\eta_j}^{2n+1} \partial_x^{2n+1}u_{0\, j}
\\ &=\eps\, (-1)^{j-1}\left( 
{\eta_j}{u}_{0\,j\,  x}-\frac{\eps^2
}{6} {\eta_j^3}\,{u}_{0\,j\,  xxx}+O(\eps^4) \right)\, \\ 
\ou_j&=\sum_{n=0}^\infty\frac{(-1)^n}{(2n+1)!}\eps^{2n}{\eta_j}^{2n}\partial_x^{2n}u_{0\, j}
= u_{0\, j}
-\frac{\eps^2}{6} {\eta_j^2}\,u_{0\,j\,  xx}+O(\eps^4)\,.
\end{split}
\end{equation}

It should be noticed that, contrary to~\cite{Wu2000,Wh2000}, for the time being we do not rescale the dependent variables $u, w$; this will be done at a later stage, when we shall rescale the Hamiltonian variable~$\sigma$ once the constraints mentioned above in Section~\ref{SSEF} will be taken into account. 
 
At leading order in the expansion with respect to the small dispersion parameter $\epsilon$, we have
\begin{equation}
\ol{u}_j=\wit{u}_j, \, \ol{w}_j=\wit{w}_j\simeq 0, \, \text{ with } \sigma=\rho_2\wit{u}_2-\rho_1\wit{u}_1=\rho_2\ol{u}_2-\rho_1\ol{u}_1,
\end{equation}
that is, $\sigma$ reduces to the {\em horizontal} momentum shear. At this order one can view the motion as satisfying the so-called {columnar} motion {\em ansatz} (see, e.g.,~\cite{PCH}).
Thus at higher orders the ansatz fails, since we have
\begin{equation}
\label{sigmaeps2}
\sigma=\rho_2\wit{u}_2-\rho_1\wit{u}_1+\eps \zeta_x(\rho_2\wit{w}_2-\rho_1\wit{w_1})
\end{equation}
and columnar motion is no longer consistent with~(\ref{uw-eps}).

For the reader's convenience, we now collect in compact form a few consequences of the expansions (\ref{intf-vel-eps}), that can be found in \S13 of \cite{Wh2000}. First, 
%
 from (\ref{intf-vel-eps}) notice that inverting
 \begin{equation}\label{u0toutilde}
\wit{u}_j=u_{0\, j}-\frac{\epsilon^2}{2} \eta_j^2 u_{0\,j\,  xx}+O(\eps^4)\, .
\end{equation}
yields
\begin{equation}
\label{utildetou0}
u_{0\, j}=\wit{u}_j+\frac{\eps^2}{2}\eta_j^2\wit{u}_{j\, xx}+O(\epsilon^4)\, .
\end{equation}
A straightforward computation shows that 
\begin{equation}\label{utildetowtilde}
\wit{w}_j=(-1)^{j+1}\, \epsilon\, \left( 
\eta_j\wit{u}_{j\, x}+\frac{\eps^2
}{3} (\eta_j ^3\, \wit{u}_{j\, xx})_x+O(\eps^4)\right)\, . 
\end{equation}
Also, as far as the asymptotic relations between interface and layer-averaged velocities are concerned, we have, again from (\ref{intf-vel-eps}), 
\begin{equation}
\label{u0toou}
\ou_j=u_{0\, j}
-\frac{\eps^2}{6} \eta_j^2\, u_{0\,j\,  xx}+O(\eps^4)\,  , 
\end{equation}
which yields,  thanks to~(\ref{utildetou0}),
\begin{equation}
\label{utildetoou}
\ou_j=\wit{u}_j+\frac{\eps^2}{3}\eta_j^2\, \wit{u}_{j\, xx}+O(\epsilon^4)\, .
\end{equation}
The mass conservation laws for the two fluids, expressed without approximation by the pair of equations
\begin{equation}
\label{exmce}
\eta_{j\,t}+\partial_x(\eta_j\,\ou_j)=0,\quad j=1,2\, ,  
\end{equation}
%
%
%
%
are translated, 
by (\ref{utildetoou}),  into the approximate mass conservation laws
\begin{equation}
\label{appmce} 
\eta_{j\,t}+\partial_x(\eta_j\,\wit{u}_j) +\frac{\epsilon^2}{3} \partial_x(\eta_j^3\wit{u}_{j \, xx})=O(\eps^4)\,, \quad \, j=1,2\,.
\end{equation}
Hence, the dynamic constraint~(\ref{dynacon}) obtained by 
summing the two equations~(\ref{exmce}), taking into account the geometric constraint $\eta_1+\eta_2=h$ together with the far-field vanishing conditions, translates  into the
%
%
%
{\em approximate dynamical constraint}
\begin{equation}\label{apprDC}
\eta_1\,\wit{u}_1+\eta_2\,\wit{u}_2+\frac{\epsilon^2}{3}\left(\eta_1^3\wit{u}_{1 \, xx}+\eta_2^3\wit{u}_{2 \, xx}\right)=O(\eps^4)\, .
\end{equation}

\subsection{The energy} 
\label{energ}
Our next task is to write the explicit form (at order $O(\eps^2)$) of the energy. 
All the asymptotic manipulations needed are for the kinetic energy,  the potential energy is straightforward and can be written out immediately at every order. \colr{The asymptotic analysis is somewhat equivalent to that used in other approaches, (see, e.g.,~\cite{CGK05}) starting from the different viewpoint of expanding, having assumed two-layer dynamics from the outset, the so-called Dirichlet-to-Neumann operator in each layer, and we can anticipate here that it will produce similar dispersive terms in the long-wave expansions below. However, besides the different starting point of geometric Hamiltonian reduction, our approach will also focus on the need to allow for different balances between nonlinearity and dispersion to capture, both qualitatively and quantitatively, fundamental features of the dynamics, while simultaneously striving for the simplest possible models.}

Let us consider the lower fluid first. Its kinetic energy density reads
\begin{equation}
\label{T2-0} 
T_2=\frac{\rho_2}{2}\, \int_{-h_2} ^\zeta(u_2^2+ w_2^2)\, h\,\D z\, ,
\end{equation}
(the dimensional factor $h$ coming from the scaling of $z$).
By Taylor-expanding, we have
\begin{equation}
u_2(x,z)=u_{2\,0}(x) -\frac{\eps^2}{2} (z+h_2)^2u_{2\,0\, xx}(x)+O(\eps^4)\, .
\end{equation}
By (\ref{utildetou0}) we get 
\begin{equation}
u_2(x,z)=\wit{u}_2(x)+\frac{\eps^2}{2}\left(\eta_2^2(x)-(z+h_2)^2\right)\wit{u}_{2\, xx} (x)+O(\eps^4)\, ,
\end{equation}
and by (\ref{uw-eps}) and (\ref{utildetou0}), 
\begin{equation}
w_2(x,z)=-\epsilon(z+h_2)\wit{u}_{2\,x}(x) +O(\eps^2).
\end{equation}
This leads to 
\begin{equation}
\label{T2}
\begin{split} 
T_2&=\frac{h\, \rho_2}{2}\int_{-h_2}^\zeta \left[ \wit{u}_2^2+\epsilon^2 \Big( \wit{u}_2\wit{u}_{2\, xx}\big(\eta_2^2-(z+h_2)^2\big)+\wit{u}_{2\,x}^2\, (z+h_2)^2\Big)
+O(\eps^4)\right] \D z \\& =
\frac{h\, \rho_2}{2}\left[\eta_2\wit{u}_2^2 +\frac{\eps^2}{3} \eta_2^3 \big(2\wit{u}_2\wit{u}_{2\,xx}+\wit{u}_{2\,x}^2\big) \right]+O(\eps^4) \, .
\end{split}
\end{equation}
By the same arguments we obtain the contribution to the total kinetic energy density of the upper fluid as
\begin{equation}
\label{T1} 
T_1=\frac{h\, \rho_2}{2}\int_\zeta^{h_1} (u_1^2+ w_1^2)\, \D z=
\frac{h\, \rho_1}{2}\left[\eta_1\wit{u}_1^2 +\frac{\eps^2}{3} \eta_2^3 \big(2\wit{u}_1\wit{u}_{1\,xx}+\wit{u}_{1\,x}^2\big) \right]+O(\eps^4) \, .
\end{equation}
In formulas (\ref{T2},\ref{T1}) we used, respectively, $\eta_2=\zeta+h_2$ and $\eta_1=h_1-\zeta$.

As mentioned above, the computation of the  potential energy density is more direct: taking non-dimensionalization into account, we have
\begin{equation}
\label{Epot}
U=h^2 \,g\left(\int_{-h_2}^\zeta\rho_2 z\, \D z+\int_{\zeta}^{h_1} \rho_1 z\, \D z \right)= 
\frac12 h^2\left( g (\rho_2-\rho_1) \zeta^2-\frac12 g(\rho_2 h_2^2-\rho_1h_1^2)\right)\, , 
\end{equation}
where $h$ is again the total distance between top and bottom plates.

\section{Nonlinear asymptotics}
\label{WNLsect}

In what follows, we shall deal with a simplified model,  defined by the following requirements:
\begin{enumerate}
\item
The interface displacement $\zeta$ will be understood to be scaled by its maximum value $a$, to yield the amplitude nondimensional small  parameter
$\alpha=\dsl{\frac{a}{h}}\ll 1$. Namely, the non-dimensional  fluid thicknesses $\eta_j$ will be written as
\begin{equation}
\label{etascaled}
\eta_j=h_j+(-1)^j \alpha\,\zeta\, .
\end{equation}
\item We shall make an asymptotic expansion in the small parameters $\alpha$ and $\epsilon$ 
and mainly consider the ``Mildly Non-Linear" (MNL) case, defined by the relative scaling
$\epsilon^2\ll \alpha\ll \epsilon$. We shall thus discard terms of order $ \alpha\epsilon^2, \epsilon^3$ and higher, but retain terms of order $\alpha^2$.
The usual Weakly Non-Linear (WNL) case (see, e.g., \cite{Wh2000}), where $\alpha=O(\epsilon^2)$, can be seen as a special case of the MNL case (see Section \ref{2Red}). 
\end{enumerate}
The first  consequences of such scaling limits are the following:
\begin{description}
\item[ i)] 
The slope $\zeta_{x}$ of the normalized interface is small and scales as $\dsl{\frac{a}{h}\frac{h}{L}} =O(\alpha\epsilon)$.
\item[ii)] Since $\wit{w}_j$ scales as $\eps$ and, by the previous point, $\zeta_{x}$ scales as $\alpha\eps$, 
the  Hamiltonian variable 
\begin{equation}
\label{sigmaB}
\sigma= \rho_2\wit{u}_2-\rho_1\wit{u}_1+\zeta_{x}(\rho_2\wit{w}_2-\rho_1\wit{w}_1)
\end{equation} 
within this asymptotics 
becomes 
\begin{equation}
\label{sigmaBexpr}
\sigma=\rho_2\wit{u}_2-\rho_1\wit{u}_1\, ,
\end{equation}
which has the same form as that of  the dispersionless approximation.
\item[iii)]
The approximate dynamical constraint (\ref{apprDC}) gets simplified as well, and reads
\begin{equation}
\label{apprDCB4}
\eta_1\wit{u}_1+\eta_2\wit{u}_2+\frac{\epsilon^2}{3}(h_1^3 \wit{u}_{1\, xx}+h_2^3\wit{u}_{2\, xx})=O(\eps^4)\, .
\end{equation}
\item[iv)] There is a notable simplification in the kinetic energy densities (\ref{T2}) and~(\ref{T1}). Indeed, for the lower fluid under approximations~(\ref{T2}) one gets
\begin{equation}
\label{T2B}
T_2=\frac{{h\, \rho_2}}{2}\left( \eta_2\wit{u}_2^2 +\frac13\eps^2 h_2^3 \Big(2\wit{u}_2\wit{u}_{2\,xx}+\wit{u}_{2\,x}^2\Big) \right)\,.
\end{equation}
\colr{
Notice that the $\epsilon^2$ term can be written as $\dsl{\frac{h_2^3}{3}} \wit{u}_2\wit{u}_{2\,xx}+\dsl{\frac{h_2^3}{6}}(\wit{u}_2^2)_{xx}$, 
and this second term, being a total derivative, does not contribute to the Hamiltonian formulation of the equations of motion}.

Repeating the argument for the upper fluid, the total kinetic energy density, still within the same 
MLN~asymptotics,  can be written as 
\begin{equation}
\label{TB}
T=T_1+T_2=\colr{
\frac{h}{2}\left( \rho_1\big(\eta_1\wit{u}_1^2+\frac{\eps^2}3\, h_1^3(\wit{u}_1\wit{u}_{1\,xx}+\frac12(\wit{u}_{1}^2)_{xx})\big)+
\rho_2\big(\eta_2\wit{u}_2^2 +\frac{\eps^2}3h_2^3(\wit{u}_2\wit{u}_{2\,xx}+\frac12(\wit{u}_{2}^2)_{xx}\big)\right)\, .}
\end{equation}
\end{description}
\subsection{The Hamiltonian in Darboux coordinates} 
Our next task is to express the Hamiltonian density $H=T+U$ of the asymptotic  model in terms of the Darboux coordinates dictated by the Hamiltonian reduction process of \S\ref{HamredMWR}, that is, the pair $\zeta$ and $\sigma$ given by  (\ref{sigmaBexpr}).
To this end, we make use of the geometrical constraint $\eta_1+\eta_2=h$ and the approximate dynamical constraint given by equation~(\ref{apprDC}).
Our strategy will be to use the weak nonlinearity assumption to simplify the dispersive terms first, and deal with small $\alpha$ expansion for the quasilinear terms afterwards, 
since the latter do not contain $x$-derivatives in the Hamiltonian, which can then be expanded in a standard Taylor series.

As remarked above, within the present asymptotic theory, the dynamical constraint reads
\begin{equation}
\label{apprDCB}
\eta_1\wit{u}_1+\eta_2\wit{u}_2+\frac{\epsilon^2}{3}(h_1^3 \wit{u}_{1\, xx}+h_2^3\wit{u}_{2\, xx})=0\, .
\end{equation}
Rewriting the latter in operator form as the equality
\begin{equation}
\label{dcop}
\eta_1\left( \mathbf{1}+\frac{\eps^2}{3}h_1^2\partial_x^2\right)\wit{u}_1=- \eta_2\left( \mathbf{1}+\frac{\eps^2}{3}h_2^2\partial_x^2\right)\wit{u}_2\, ,
\end{equation}
which is correct up to terms of order $\alpha\, \epsilon^2$, and
by using the approximate inversion formula for near-identity operators
$(\mathbf{1}+\eps^2 \widehat{A})^{-1}=\mathbf{1}-\eps^2 \widehat{A}+O(\eps^4)$,
we get
\begin{equation}
\label{dcop2}
\wit{u}_1=-\left(\mathbf{1}-\frac{\eps^2}3 h_1^2\partial_x^2 \right)\left(\frac{\eta_2}{\eta_1}\big( \mathbf{1}+\frac{\eps^2}3 h_2^2\partial_x^2\big)\right)\wit{u}_2\, ,
\end{equation}
up to higher order terms in $\eps^2$.
Since $\dsl{\frac{\eta_2}{\eta_1}}=\dsl{\frac{h_2}{h_1}}+O(\alpha)$ we arrive at the  relation
\begin{equation}
\label{u2tou1}
\wit{u}_1=-\frac{\eta_2}{\eta_1}\wit{u}_2+\frac{\eps^2}{3}\frac{h_2}{h_1}(h_1^2-h_2^2)\wit{u}_{2\, xx}\, .
\end{equation}
{\color{black} 
Recall that the kinetic energy density is represented, at $O(\eps^2)$ and in this weakly non-linear asymptotics, by 
\begin{equation}
\label{TB2}
{T}=
\frac{h}{2}\left( \rho_1\big(\eta_1\wit{u}_1^2+\frac{\eps^2}3\, h_1^3\wit{u}_1\wit{u}_{1\,xx}\big)+
\rho_2\big(\eta_2\wit{u}_2^2 +\frac{\eps^2}3h_2^3\wit{u}_2\wit{u}_{2\,xx}\big)\right)\, \colr{\text{plus total derivatives}.}
\end{equation}
At the order of approximation we are working with we can substitute
\begin{equation}
\label{u1xx}
\wit{u}_{1\,xx} =-\frac{h_2}{h_1} \wit{u}_{2\,xx}
\end{equation}
in the $O(\epsilon^2)$ terms of this expression as well as in the approximate dynamical constraint (\ref{apprDCB}), which therefore turns into
\begin{equation}\label{naDC}
\wit{u}_{{1}}\eta_{{1}}+\wit{u}_{{2}}\eta_{{2}}+\frac13\,{\epsilon}^{2} \wit{u}_{{2\,{xx}}}h_2(h_2^2-h_1^2)=0\, .
\end{equation}
By solving this algebraic constraint and the defining relation (\ref{sigmaBexpr}) with respect to the velocities, we get the implicit relations
\begin{equation}
\label{u-s-uxx}
\begin{split}
\wit{u}_1&=-{\frac {\sigma\,\eta_{{2}}}{\eta_{{1}}\rho_{{2}}+\eta_{{2
}}\rho_{{1}}}}+\frac{{\epsilon}^{2}}3\,{\frac {\wit{u}_{{2\,{\it xx}}}h_{{2}}\rho_{{2}} \left( {h_{{1}}}^{2}-{h
_{{2}}}^{2} \right)}{\eta_{{1}}\rho_{{2}}+\eta_{{2}}
\rho_{{1}}}}\\
\wit{u}_2&=\frac {\sigma\,\eta_{{1}}}{\eta_{{1}}\rho_{{2}}+\eta_{{2
}}\rho_{{1}}}
+\frac{{\epsilon}^{2}}3\,\frac {\wit{u}_{{2\,{\it xx}}}h_{{2}}\rho_{{1}} \left( {h_{{1}}}^{2}-{h
_{{2}}}^{2} \right) }{{\eta_{{1}}\rho_{{2}}+\eta_{{2}}
\rho_{{1}} }}
\end{split}
\end{equation}
Now we can use the fact that the second derivative $\wit{u}_{2\, xx}$ appears only in terms $O(\eps^2)$, so that we can substitute~(\ref{u1xx}) into 
\begin{equation} 
\sigma_{{{\it xx}}}=\rho_{{2}}\, \wit{u}_{2\,{{\it xx}}}-\rho_{{1}}\, \wit{u}_{1\,{{\it xx}}}
\label{sigint}
\end{equation} 
leading to 
\begin{equation}
\wit{u}_{2\,{{\it xx}}}={\frac {\sigma_{{{\it xx}}}h_{{1}}}{h_{{1}}
\rho_{{2}}+h_{{2}}\rho_{{1}}}}\, .
\end{equation}
Hence, equations~(\ref{u-s-uxx}) become
\begin{equation}
\label{u-s-sxx}
\begin{split}
\wit{u}_1&=-{\frac {\eta_{2} \,\sigma}{\eta_{{1}}\rho_{{2}}+\eta_{{2
}}\rho_{{1}}}}+ \rho_{{2}}\frac{{\epsilon}^{2}}3\,{\frac {h_1\, h_{{2}}  \left( {h_{{1}}}^{2}-{h
_{{2}}}^{2} \right)}{(h_{{1}}\rho_{{2}}+h_{{2}}\rho_{{1}})({\eta_{{1}}\rho_{{2}}+\eta_{{2}}\rho_{{1}} )}}\sigma_{xx}
}\,,
\\
\wit{u}_2&=\frac {\eta_{1}\,\sigma}{\eta_{{1}}\rho_{{2}}+\eta_{{2
}}\rho_{{1}}}
+\rho_{{1}}\frac{{\epsilon}^{2}}3\,\frac {h_1\, h_{{2}} \left( {h_{{1}}}^{2}-{h_{{2}}}^{2} \right) }
{(h_{{1}}\rho_{{2}}+h_{{2}}\rho_{{1}})({\eta_{{1}}\rho_{{2}}+\eta_{{2}}\rho_{{1}} )}}\sigma_{xx}\,.
\end{split}
\end{equation}
Substituting these relations in the expression of the kinetic energy density (\ref{TB2}) leads, \colr{dropping the total derivative terms,}  to the intermediate expression
\begin{equation}
\label{T2s-eta}
{T}=\frac{h}2 \left(\frac{\eta_1\eta_2 \sigma^2}{\rho_2\eta_1+\rho_1\eta_2}+ \frac{\epsilon^2}3{\frac {{h_{{1}}}^{2}{h_{{2}}}^{2}
 \left( h_{{1}}\rho_{{1}}+h_{{2}}\rho_{{2}} \right) }{ \left( h_{{1}}
\rho_{{2}}+h_{{2}}\rho_{{1}} \right) ^{2}}} \sigma\,\sigma_{{{\it xx}}}\right)\,.
\end{equation}
Next, the first term in the kinetic energy must be expanded in powers of $\alpha$ to yield our final version of the kinetic energy density
\begin{equation}
\label{Talpha}
\begin{split} 
{T}=&
\frac{h}2\frac{h_1\, h_2}{\left( h_1
\rho_2+h_2\rho_1 \right)}\sigma^2+\frac{\alpha}{2} \frac{h\,(h_1^2\rho_2-h_2^2\rho_1)}{{\linden}^2}\, \zeta \, \sigma^2-
\frac{\alpha^2}2\frac{\colr{h^3}\,\rho_1\rho_2}{\linden^3}\,\zeta^2 \sigma^2 
\\
& + \frac{\epsilon^2}6 {\frac {h\, {h_{{1}}}^{2}{h_{{2}}}^{2}
 \left( h_{{1}}\rho_{{1}}+h_{{2}}\rho_{{2}} \right) }{ \left( h_{{1}}
\rho_{{2}}+h_{{2}}\rho_{{1}} \right) ^{2}}}\, \sigma \, \sigma_{{{\it xx}}}
+O(\alpha^3, \alpha\eps^2,\eps^4)\, .
\end{split}
\end{equation}
Therefore,  with the potential energy expression~(\ref{Epot}), the total energy density at this order  is
\begin{equation}
\label{Etot}
{
\begin{split} 
{E}=&
h\left(\frac{1}2\frac{h_1\, h_2}{\left( h_1
\rho_2+h_2\rho_1 \right)}\sigma^2+\frac{\alpha}{2} \frac{(h_1^2\rho_2-h_2^2\rho_1)}{{\linden}^2} \zeta\sigma^2-
\frac{\alpha^2}2\frac{h^2\rho_1\rho_2}{\linden^3}\zeta^2\, \sigma^2\right)
\\
& + h
\frac{\epsilon^2}6 {\frac {{h_{{1}}}^{2}{h_{{2}}}^{2}
 \left( h_{{1}}\rho_{{1}}+h_{{2}}\rho_{{2}} \right) }{ \left( h_{{1}}
\rho_{{2}}+h_{{2}}\rho_{{1}} \right) ^{2}}} \sigma\,\sigma_{{{\it xx}}} 
+\frac12 h^2\, g (\rho_2-\rho_1)  \zeta^2\, .
\end{split}}
\end{equation}
It is convenient to introduce the non-dimensional momentum shear 
$\sigma^*$ 
by
\begin{equation}
\label{astvar}
\sigma=\sqrt{h\, g (\rho_2-\rho_1)\linden}\>\>\sigma^*\, 
\end{equation}
so that the non-dimensional form \colr{$E^*$} of the total energy is (immediately dropping asterisks for ease of notation)
\begin{equation}
\label{ndEn}
\begin{split}
E&=
\frac{1}2{h_1\, h_2}{\sigma}^2+
\frac{\alpha}{2} \frac{{h_1}^2\rho_2-{h_2}^2\rho_1}{{\linden}}\, \zeta\, {\sigma}^2
-\frac{\alpha^2}2\frac{\colr{h^2}\, \rho_1\rho_2}{\linden^2}\, {\zeta}^2\, {\sigma}^2
\\
& 
\ \ \ + \frac{\epsilon^2}6 {\frac {{h_{{1}}}^{2}{h_{{2}}}^{2}
 \left( h_{{1}}\rho_{{1}}+h_{{2}}\rho_{{2}} \right) }{ h_{{1}}
\rho_{{2}}+h_{{2}}\rho_{{1}} }}\, \sigma\,\sigma_{{{\it xx}}} +\frac12 {\zeta}^2\, 
\\ &=
\frac12\left(A\, {\sigma}^2+\alpha B\,  \zeta{\sigma}^2-\alpha^2 C\,  {\zeta}^2{\sigma}^2+{\zeta}^2+\eps^2 \kappa\sigma\,\sigma_{{{\it xx}}} \right)
\end{split}
\end{equation}
where 
we denoted the constants by 
\begin{equation}
\label{ABC}
A=h_1\, h_2,\quad B= \frac{{h_1}^2\rho_2-{h_2}^2\rho_1}{{\linden}}, \quad C
=\frac{\colr{h^2}\rho_1\rho_2}{\linden^2},\quad\kappa={ \frac13}{\, \frac {{h_{{1}}}^{2}{h_{{2}}}^{2}
 \left( h_{{1}}\rho_{{1}}+h_{{2}}\rho_{{2}} \right) }{ h_{{1}}
\rho_{{2}}+h_{{2}}\rho_{{1}} }}\, .
\end{equation}
Applying the Poisson tensor (\ref{Pred}) to the variational differential of the energy (in fact, the Hamiltonian)  $\mathcal{E}=\int E \, \D x$ yields the equations of motion as 
\begin{equation}
\label{EMformal}
\left(\begin{array}{c}\medskip \zeta_t\\ \sigma_t\end{array}\right)=-\left(\begin{array}{cc}\medskip
0&\partial_x\\
\partial_x&0\end{array}\right) \, \left(\begin{array}{c}\medskip \displaystyle\frac{\delta \mathcal{E}}{\delta \zeta} \\ \displaystyle\frac{\delta \mathcal{E}} { \delta\sigma}\end{array}\right)
\end{equation}
where $t$ is the non-dimensional time, related with the physical time by
\begin{equation}
t\to \eps\sqrt{\frac{g(\rho_2-\rho_1)}{h\, \linden}}\,\, t\, .
\end{equation}
This shows explicitly how the evolution proceeds in a slow time gauged by the dispersion parameter $\eps$ as required by the long-wave asymptotics.

The resulting system in conservation form is
\begin{equation}
\label{eqofmot}
\left\{\begin{split}
&\zeta_t+\left(A {\sigma}+\alpha B{\zeta}{\sigma}-\alpha^2 C{\zeta}^2{\sigma}+\eps^2\kappa{\sigma}_{xx}\right)_x=0\\
&\sigma_t+\left({\zeta}+\alpha \frac{B \,{\sigma}^2}{2} -\alpha^2 C{\zeta}{\sigma}^2\right)_x=0\end{split}\right.\, , 
\end{equation}
or, carrying out the relevant spatial differentiations  explicitly,
\begin{equation}
\left\{\begin{split}
&\zeta_t+A \sigma_x+\alpha B({\zeta}{\sigma})_x-\alpha^2 C({\zeta}^2{\sigma})_x+\eps^2 \kappa{\sigma}_{xxx}=0
\\
&\sigma_t+\zeta_x+\alpha B {\sigma}{\sigma}_x-\alpha^2 C({\zeta}{\sigma}^2)_x=0
\end{split}\right.\, . 
\label{exeom}
\end{equation}
which from now on will be referred to as the {\em ABC-system}. 

\remark
\label{rmk:phys-par}
\rm
A few comments on the parameters $A,B,C$ and their relations with the physical parameters $\rho_1, \rho_1, h_1, h_2$ are in order. First, the parameter $A$ is just the square of the linear wave velocity; in nondimensional form it ranges from $0$ to $\frac14$, and could be set to unity by further rescaling  $\sigma$. Next, note that 
the parameter $\kappa$ is nonnegative, and vanishes only when $h_1\to 0$ or $h_2\to 0$.
Similarly, the parameter $C$ is nonnegative, and vanishes only in the air-water limit $\rho_1\to 0$.
The most interesting parameter is $B$, which is non sign-definite and appears in front of the cubic term $\sigma^2\zeta$ of the Hamiltonian. It vanishes at the critical ratio
\begin{equation}
\label{critval}
\frac{\rho_1}{\rho_2}=\frac{h_1^2}{h_2^2}\,.
\end{equation}
By denoting  the density ratio parameter $r=\dsl{{\rho_1}/{\rho_2}}$,  so that $0<r<1$, 
the definition of $B$ shows that it is positive for $h_1>\sqrt{r}\, h_2$, and negative for $h_1< \sqrt{r} \, h_2$. One of the most relevant effects of this change in sign shows up in the existence and polarity of solitary travelling wave solutions of system~(\ref{eqofmot}), as we shall see below in Section~\ref{specsol}. In this regard,  to account for the break down of the theory near the vanishing of the $B$ coefficient of quadratic nonlinearity, we notice that under the WNL asymptotic scaling it would be necessary to compute a plethora of higher order terms for asymptotic consistency, as shown in~\cite{Nguyen-Dias}. Such terms involve higher order derivatives which make the search for travelling solutions a (mostly) numerical affair, whereas reasonable qualitative and somewhat quantitative agreement with Euler solutions can already be obtained under the present ABC model.

\remark \rm
The Boussinesq approximation consists of retaining density differences in the potential (gravitational) energy density, while neglecting the associated inertial differences in the kinetic energy density,  by setting in~(\ref{TB})
\begin{equation}
\label{rhobar} 
\rho_1=\rho_2=\bar{\rho}\, .
\end{equation}
This Boussinesq approximation simplifies significantly the weakly or mildly nonlinear asymptotics; indeed, the Hamiltonian variable for the weighted shear reduces to $\sigma=\bar{\rho}(\wit{u}_2-\wit{u}_1)$, and the non-dimensional energy density of the system becomes
\begin{equation}
\label{Hred} 
\begin{split}
{E}_B&=
\frac12{h_1\, h_2}{\sigma}^2+
\frac{\alpha}{2} (h_1-h_2) \zeta{\sigma}^2
-\frac{\alpha^2}2{\zeta}^2\, {\sigma}^2
+ \frac{\epsilon^2}6 {h_1}^{2}{h_{{2}}}^{2} \sigma\,\sigma_{{{\it xx}}} +\frac12 {\zeta}^2\, \\ &=
\frac12\left(A_B\, {\sigma}^2+\alpha B_B\,  \zeta{\sigma}^2-\alpha^2 C_B\,  {\zeta}^2{\sigma}^2+{\zeta}^2+\eps^2\, \kappa_B\sigma\,\sigma_{{{\it xx}}} \right),
\end{split}
\end{equation}
where 
\begin{equation}
\label{ABC-Bouss}
A_B=h_1\, h_2,\quad B_B= h_1-h_2, \quad C_B=1,\quad\kappa_B=\frac13 \,{h_1}^{2}{h_2}^{2}\, .
\end{equation}

}

The Hamiltonian formulation of system~(\ref{eqofmot}) provides three additional constants of the motion besides the energy $\mathcal{E}$. They are the two Casimir functionals, 
\begin{equation}
\label{Kaz}
\mathcal{K}_1=\int_\RR \zeta\, \D x,\quad \mathcal{K}_2=\int_\RR \sigma\, \D x, 
\end{equation}
and the generator of the $x$-translation
\begin{equation}
\label{M-I} 
\Pi=\int_\RR \zeta\, \sigma\, \D x\,.
\end{equation}
They are conserved quantities for any choice of the parameters $A,B$ and $C$. As we shall see in the following section, the weakly nonlinear case  is rather special, as it reduces to the so-called completely integrable Kaup-Boussinesq systems.  

\section{Two notable reductions and their complete integrability}
\label{2Red}

It is worth considering further simplifications of the reduction~(\ref{EMformal}) as they may be applicable to certain physical regimes and offer the unexpected bonus of being completely integrable. We first 
look at the weakly nonlinear limit (WNL) in the context of the bidirectional system~(\ref{eqofmot}). We then examine how the Hamiltonian reduction strategy can be used to derive unidirectional motion equations.

\subsection{The WNL case and the Kaup-Boussinesq system}

The WNL asymptotics mentioned above, where $\alpha=O(\eps^2)$, formally corresponds to dropping the $C$-term in the Hamiltonian, as the quartic terms of $O(\alpha^2)$ becomes subdominant with respect to other terms unless the ``hardware" parameters (depths and densities) are near the critical ratio (\ref{critval}), where the coefficient of the cubic term $B$ vanishes. Away from the critical ratio, the WNL case leads to a ``universal" representative bidirectional system which can be viewed as standing at the same level as its unidirectional counterpart, the well known KdV equation.

Suppressing the order parameters $\alpha$ and $\eps$ for ease of notation, system~(\ref{exeom}) in the WNL limit becomes
\begin{equation}
\label{kaupsys}
\left\{\begin{split}
&\zeta_t+A \sigma_x+ B(\zeta\sigma)_x+ \kappa\sigma_{xxx}=0
\\
&\sigma_t+\zeta_x+ B \sigma\sigma_x=0
\end{split}\right.\, .
\end{equation}
This system is a parametric version of the Boussinesq system for water waves equations, introduced by \cite{Broer75,Wh2000}. It is asymptotically equivalent, up to terms of order $O(\alpha,\epsilon^2)$, to the nonlocal one reported in \cite{CC99} (the form apparently favored by Boussinesq~\cite{Wh2000},\S13.11), through the change of variables 
\begin{equation}
\label{sigbar}
\bar{\sigma}= \sigma +\kappa\sigma_{xx}\,,\quad \hbox{and}\quad \bar{\sigma}=\left({\rho_1h_2+\rho_2 h_1 \over h_2}-{\rho_2\zeta \over h_2^2}\right) \bar{u}_1\,.
\end{equation} 
System~(\ref{kaupsys}) is completely integrable via the Inverse Scattering Method, as shown in~\cite{Kaup75}, and further analyzed in~\cite{Ku85}, where it was also shown how it can be derived in bi-Hamiltonian form. In our variables (which are related to those in~\cite{Ku85} by a nontrivial Miura-like transformation) the corresponding Poisson pencil (see, e.g.,~\cite{DubZha2010,FMP98}) is
\begin{equation}
\label{poi-pen}
P(\la)=\la P_0-P_1=\begin{pmatrix} \frac12 B(\zeta\del_x+\del_x\zeta)+A\del_x+
\kappa\del_x^3 & 
\left(\frac12 B\sigma-\lambda\right)\del_x\\
 & \\
\del_x \left(\frac12 B\sigma-\lambda\right) & \del_x
\end{pmatrix}\,.
\end{equation}
Indeed, the equations of motion (\ref{kaupsys}) can be written as the Hamiltonian evolution 
\begin{equation}
\label {bHform}
\begin{pmatrix} \medskip\zeta_t\\ \sigma_t\end{pmatrix}=P_1\begin{pmatrix}\medskip \displaystyle\frac{\delta \Pi}{\delta\zeta}\\  \displaystyle\frac{\delta \Pi}{\delta\sigma}\end{pmatrix} = P_0\begin{pmatrix}\medskip \displaystyle\frac{\delta \mathcal E}{\delta\zeta}\\  \displaystyle\frac{\delta \mathcal E}{\delta\sigma}\end{pmatrix}\, .
\end{equation}
\colr{Throughout this section and the next one, the differential operator $\partial_x$ is, as usual, meant to act on all quantities that stand to its right, e.g.,  $\partial_x\, \zeta\phi=(\zeta\,\phi)_x$. Also, in the above formula,  
$\Pi$ is the generator of $x$-translations (\ref{M-I}) and we renamed $P_0$ the tensor $J_{\rm red}$ of (\ref{Pred}).}\\
This bi-Hamiltonian formulation can be used to construct recursively an infinite family of constants of  motion.
We briefly review  here the technique 
in  \cite{FMP98}, adapted to system~(\ref{kaupsys}). First, we seek the Casimir of the Poisson pencil (\ref{poi-pen}), in the form of 
a series $\mathcal{H}(\la)$ in inverse powers of $\lambda$, $\mathcal{H}(\la)=\sum_{n=0}^\infty{\mathcal H}_n \la^{-n}$,
whose variational gradient satisfies
\begin{equation}
\label{CPP1}
P(\la) \cdot d\, \mathcal{H}(\la)=0\, .
\end{equation}
Denoting by $(\gamma, \beta)$ the components of the gradient of $\mathcal{H}(\la)$ one gets the following system
\begin{equation}
\label{cas1}
\left\{
\begin{array}{l}\medskip
\frac{B}2(\zeta\gamma_x+(\zeta\gamma)_x)+A\gamma_x+\kappa\gamma_{xxx}+(\frac{B}2\sigma-\la) \beta_x=0\\
\frac{B}2(\sigma\gamma)_x-\la\gamma_x+\beta_x=0
\end{array}
\right.
\end{equation}
Substituting $\beta_x=\la\gamma_x-\dsl B (\sigma\gamma)_x/2$ from the second equation into the first 
yields an expression for $\gamma$ that can be manipulated, by multiplying  it by $\gamma$, into the total $x$-derivative. 
\begin{equation}
\label{Cas2}
\frac{B}2[(\gamma\zeta)\gamma_x+\gamma(\gamma\zeta)_x]-\frac{B^2}{4}(\gamma\sigma)(\gamma\sigma)_x+\la\frac{B}{2}[(\gamma\sigma)\gamma_x+\gamma(\gamma\sigma)_x]+(A-\la^2)\gamma\gamma_x +\kappa \gamma\gamma_{xxx}=0\,.
\end{equation}
By integrating in~$x$, system~(\ref{cas1}) can be replaced by
\begin{equation}
\label{Cas3}
\left\{
\begin{array}{l}\medskip
\frac12\left(-\la^2+A+B(\la\sigma+\zeta-\frac{B}4\sigma^2)\right)\gamma^2+\kappa(\gamma\gamma_{xx}-\frac12 \gamma_x^2)=F(\la)
\\
\frac{B}2\sigma\gamma-\la\gamma+\beta=G(\la) 
\end{array}\,,
\right.
\end{equation}
where $F(\lambda)$ and $G(\lambda)$ are the arbitrary constants of integration with respect to ${x}$.
The corresponding inverse power series for $\gamma=1+O(\frac{1}{\la})$ and  $\beta=O(\frac{1}{\la})$  can be 
obtained by setting 
\begin{equation}
F(\la)=-\frac{\la^2}{2}, \quad G(\la)=-\la\,.
\label{FG}
\end{equation}
It is straightforward to check that with this choice system~(\ref{Cas3}) can be solved iteratively. It remains to show that the one-form~$(\gamma, \beta)$ is exact. 
To this end we define
\begin{equation}
\label{hacca}
h(\la):=\frac{\la}{2\sqrt{\kappa}}\frac1{\gamma}+\frac{\gamma_x}{2\, \gamma}=\frac\la{2\sqrt{\kappa}}+h_0+\frac{h_1}{\la}+
\frac{h_2}{\la^2}+\frac{h_3}{\la^3}\cdots \, .
\end{equation}
In terms of $h(\la)$ we can write (\ref{Cas3}), subject to the choice~(\ref{FG}),  as
\begin{equation}
\label{Cas4}
\left\{
\begin{array}{l}\medskip
h(\la)_x+h(\la)^2=\dsl{\frac1{4\kappa}}\left( \la^2- A-B\left(\la \sigma+\zeta-\dsl{\frac{B\sigma^2}4}\right)\right)
\\
\beta(\la)=\la(\gamma-1)-{\frac{B}2}\sigma\gamma\, .
\end{array}
\right.
\end{equation}
Let us consider the one-form $(\gamma, \beta_\gamma)$ with $\beta_\gamma$ given by the second equation of this system, and denote by $(\dot{\zeta},\dot\sigma) $ the tangent vector to a generic curve in the phase space~$(\zeta,\sigma)$. Then
\begin{equation}
\int_\RR(\gamma \dot\zeta+\beta_\gamma\dot{\sigma})\, \D x=\int_\RR 
\left(\gamma\dot\zeta+\left(\la(\gamma-1)-\frac{B}2\gamma\sigma\right)\dot\sigma\right)\, \D x\, .
\end{equation}
From the first of (\ref{Cas4}) we get
\begin{equation}
\label{ricdot}
\dot{h}_x+2h\dot{h}=-\frac{B}{4\kappa}\left(\dot\zeta+\la\dot\sigma-\frac{B}2 \sigma\dot\sigma \right)\, .
\end{equation}
Multiplying by $\gamma$, integrating by parts, and using the definition (\ref{hacca}) finally yields
\begin{equation}
\label{Casfin}
\int_\RR\left(\gamma \dot\zeta+\beta_\gamma\dot{\sigma}\right)\, \D x=-\la \frac{d}{dt}\int_\RR \big(\sigma+
\frac{4\sqrt{\kappa}}{B} h(\lambda)\big)\, \D x\, .
\end{equation}
We can conclude that 
\begin{equation}
\mathcal{H}(\la)=- \int_\RR \Big(\sigma+
\frac{4\sqrt{\kappa}}{B} h(\lambda)\Big)\, \D x
\label{quq}
\end{equation}
is a Casimir of the Poisson pencil (\ref{poi-pen}); hence the coefficients of its expansion in inverse powers of $\la$ are mutually 
commuting constants of the motion.
The first conserved quantities are 
\begin{equation}
\begin{split}
\mathcal{H}_1=&\int_\RR \zeta \, \D x \, , 
\\
\mathcal{H}_2=&\frac{B} 2\int_\RR   \zeta \sigma\, \D x\,  ,\\
\mathcal{H}_3=& \frac{B}2\int_\RR
\left(\frac{1}{2} A \sigma^2+\frac{1}{2} B \zeta  \sigma^2+\frac{\zeta^2}{2}-\frac{1}{2} \kappa  \sigma_x^2\,  \right)\D x\, ,
\\
\mathcal{H}_4=&  \frac{B^2}{8} \int_\RR \left(  A \sigma ^3
+B \zeta  \sigma ^3+3  \zeta ^2 \sigma -3  \kappa  \sigma  \sigma_x^2-\frac4{B} \kappa  \zeta_x \sigma_x \right)\D x\, .
\label{AB0-cons-qty}
\end{split}
\end{equation}
Note that $\mathcal{H}_1$ is a Casimir of $P_0$, the quantity $\mathcal{H}_2$ was already identified with  the generator of $x$-translations, while
 $\mathcal{H}_3$ is (up to a factor $1/2$)  the energy  (i.e., the Hamiltonian functional~(\ref{ndEn}) for $P_0$). 
 Together with $\int_{\ssbarr} \sigma\, \D x$, the first three conserved quantities come from basic physical principles. The fourth,  
 $\mathcal{H}_4$, and all the higher order $\mathcal{H}$'s thus constructed are 
the conserved quantity more directly associated with the Liouville integrability of the mathematical problem and the bi-Hamiltonian formulation we have  described. 
\colr{It is well known (see, e.g., \cite{Broer75}), that the energy $\mathcal{H}_3$ failing to be a  positive-definite quantity implies that the corresponding equations of motion are not ``well protected against short wave instability," the so-called {\em bad\/} Boussinesq equation being possibly the 
prototypical example of an integrable equation displaying such a drawback. Further comments on this phenomenon can be found in Section \ref{displin}.}

We remark that the full ABC system~(\ref{eqofmot}),  unlike its WNL reduced case,  seems to fail the complete integrability property of a second local Hamiltonian structure. Following the WNL structure, one could  make use of the conserved quantity 
(proportional to)  $\mathcal{H}_2$ above to provide such a structure with the anti-symmetric operator 
\begin{equation}	
\label{PABC}
P_{ABC}=
\begin{pmatrix} -\frac12 B(\zeta\del_x+\del_x\zeta)+A\del_x+
\kappa\del_x^3 & 
-\frac12 B\sigma\del_x +C\del_x \sigma \zeta \\
 & \\
-\frac12 \del_x  B\sigma+C\sigma \zeta \del_x  & \del_x+C \sigma\del_x \sigma 
\end{pmatrix}\,.
\end{equation}
Used with the appropriate factor of~$\mathcal{H}_2$, this operator  does yield the equations of motion~(\ref{eqofmot}); however, because its dispersionless limit is not associated with a flat metric, as detailed in~\cite{DubrovinNovikov}, $P_{ABC}$ fails to satisfy a necessary condition for fulfilling Jacobi identity, and hence cannot be used to generate a second Hamiltonian structure for system~(\ref{eqofmot}).

\subsection{Unidirectional models}
To obtain unidirectional nonlinear wave equations for our model, we at first observe that the rescaling
$\sigma\to \sqrt{A}\,\sigma$ simplifies the Hamiltonian density (\ref{ndEn}) to 
\begin{equation}
\label{Hne}
\widetilde{\mathcal{H}}=\frac12 \left({\sigma}^2+{\zeta}^2+\alpha \wit{B}\,  \zeta{\sigma}^2-\alpha^2 \wit{C}\,  {\zeta}^2{\sigma}^2+\eps^2 \wit{\kappa}\sigma\,\sigma_{{{\it xx}}}\right)
\end{equation}
(with $\wit{B}=\dsl{\frac{B}{A}}$ and so on and so forth), and the ensuing Hamiltonian equations of motion to
\begin{equation}
\left\{\begin{split}
&\zeta_t=-(\sigma_x+\alpha \wit{B}({\zeta}{\sigma})_x-\alpha^2 \wit{C}({\zeta}^2{\sigma})_x+\eps^2 \wit{\kappa}{\sigma}_{xxx})
\\
&\sigma_t=-(\zeta_x+\alpha \wit{B} {\sigma}{\sigma}_x-\alpha^2 \wit{C}({\zeta}{\sigma}^2)_x) 
\end{split}\right.\, .
\label{exeom2}
\end{equation}
We seek (following, e.g., the classical steps of \cite{Wh2000} \S 13) for a relation $\sigma=\sigma(\zeta)$ of the form
\begin{equation}
\label{1dfrel}
\sigma=\zeta+\alpha F(\zeta)+\alpha^2G(\zeta)+\epsilon^2K(\zeta)\, 
\end{equation}
with $F,G,K$ differential polynomials in $\zeta$ such that  the resulting equations obtained substituting (\ref{1dfrel}) in (\ref{exeom2}) coincide \colr{up to terms vanishing faster than $\alpha^2$ and $\epsilon^2$ in the limit $\alpha,\epsilon\to 0$, that is, at order $O(\alpha^2, \eps^2)$.}
This procedure can be  carried on in a straightforward manner,  the only difference with the derivation of the KdV equation of \cite{Wh2000} being that at order $O(\alpha)$ one has to use the relation
\begin{equation}
\label{tx-alpha}
\partial_t
=-{{\partial_x}}\,-\frac32\,\wit{B}\alpha\,\zeta    {{\partial_x}}
  \, .
 \end{equation}
The outcome is the following:
\begin{description}
\item[ i)] The link between $\zeta$ and $\sigma$ of equation (\ref{1dfrel}) is explicitly given by
\begin{equation}
\label{constr1d}
\sigma=\zeta-\frac14\,\alpha\,\wit{B}\,{\zeta}^{2}+\frac18\,{\alpha}^{2}{\wit{B}}^{
2}{\zeta}^{3}-\frac12\,{\epsilon}^{2}\wit{\kappa}\,\zeta_{{{\it xx}}}\,.
\end{equation}
\item[ii)]
The resulting unidirectional equation of motion is \colr{a parametric form of} the (defocusing) Gardner (or KdV-mKdV) equation
\begin{equation}
\label{1direq}
\zeta_{{t}}=-\zeta_{{x}}-\frac32\,\alpha \wit{B}\zeta\,\zeta_{{x}}+\left( 3\,{\alpha}^{2}\wit{C}+\frac38\,{\alpha}^{2}{\wit{B}}^{2} \right)\,{\zeta}^{2} \zeta_{{x}}
-\frac12\,{\epsilon}^{2}\wit{\kappa}\,\zeta_{{\it xxx}} \, ,
\end{equation} 
\end{description}
\colr{which  was derived in the theory of stratified fluids, e.g., in  \cite{DjRe78}.}

We first notice that, in the Weakly Non Linear (WNL) approximation, that is at $O(\alpha,\eps^2)$ with 
$\alpha=O(\eps^2)$, equation~(\ref{1direq}) becomes the Korteweg-deVries (KdV) equation, 
and relation~(\ref{constr1d}) reduces to the one of~\cite{Wh2000} \S 13. 
Moreover, although for $\wit{C}=0$ the Hamiltonian (\ref{Hne}) becomes the Hamiltonian of the Kaup-Boussinesq system (\ref{kaupsys}), the resulting unidirectional equation for $\zeta$ has a modified KdV (mKdV) term, given by $\dsl{3\,{\alpha}^{2}{\wit{B}}^{2} \,{\zeta}^{2} \zeta_{{x}}/8}$.

These unidirectional equations can be given a Hamiltonian interpretation by providing an alternative strategy by a geometric reshaping of the argument in~\cite{Olv84a,Olv84b} \colr{(which refers to a single layer Euler fluid,  an inessential difference in this context)}.
We regard equation (\ref{constr1d}) as an {\em asymptotic constraint} between the two dependent variables $\sigma, \zeta$ and we apply the Dirac theory of constraints~\cite{Dirac}, and its related  Dirac Poisson brackets.  First, a straightforward computation shows that, still in the $O(\epsilon^2,\alpha^2)$ asymptotics, no secondary constraint arise, that is, if we denote by  
\begin{equation}
\label{phidef}
\Phi\equiv \sigma-\zeta+\frac14\,\alpha\,\wit{B}\,{\zeta}^{2}-\frac18\,{\alpha}^{2}{\wit{B}}^{
2}{\zeta}^{3}+\frac12\,{\epsilon}^{2}\wit{\kappa}\,\zeta_{{{\it xx}}}=0
\end{equation}
the constraint, the equations of motion, the constraint equation~(\ref{constr1d})  and relation~(\ref{tx-alpha}) imply
\begin{equation}
\label{noseccon}
\Phi_t\approx 0\text{ at}\, O(\alpha^2, \epsilon^2)\, , 
\end{equation}
where the ``$\approx$"symbol, as per the usual Dirac's theory notation, stands for equality on the constrained manifold. 
Second,  we notice that the pair $\zeta, \varphi=\sigma-g(\zeta)$, where $g(\zeta)= \zeta-\frac14\,\alpha\,\wit{B}\,{\zeta}^{2}+\frac18\,{\alpha}^{2}{\wit{B}}^{
2}{\zeta}^{3}-\frac12\,{\epsilon}^{2}\wit{\kappa}\,\zeta_{{{\it xx}}}$, is a set of coordinates equivalent to the pair $\zeta, \sigma$, and we express the Poisson tensor (\ref{Pred}) in these new coordinates. The result is the matrix of differential operators
\begin{equation}
\label{Predtil}
\wit{P}=\left(
\begin{array}{cc} 0&-\partial_x\\
-\partial_x&\partial_x \cdot g^\prime(\zeta)+g^\prime(\zeta)\cdot \partial_x\end{array}
\right)\,,
\end{equation}
where we denoted by $g^\prime(\zeta)$ the Fr\'echet derivative of $g(\zeta)$, viz.
\begin{equation}
\label{g-prime}
g^\prime=1-\frac12 \wit{B}\alpha \zeta+\frac38\, \wit{B}^2\alpha^2\, \zeta^2-\frac12\epsilon^2\wit{\kappa}\partial_{xx}\, .
\end{equation}
In analogy with the usual formula of the Dirac Poisson brackets for the finite $N$-dimensional case $q_1, \dots q_N$, with a number $M<N$ of constraints	
$\Phi_1,\dots\Phi_M$,
\begin{equation}
\label{DirFD}
\{q_i, q_j\}^D=\{q_i, q_j\}-\sum_{a,b=1}^M \{q_i, \Phi_a\}(\mathcal{C}^{-1})_{ab} \{\Phi_b, q_j\}\, , 
\end{equation}
where $\mathcal{C}$ is the matrix with entries $\{\Phi_a, \Phi_b\}$,
the Dirac 
tensor in the coordinates $(\zeta, \varphi)$ is given  by 
\begin{equation}
\label{PDirac}
P^D\equiv\left(
\begin{array}{cc} \wit{P}_{11}-\wit{P}_{12}\left(\wit{P}_{22}\right)^{-1}\wit{P}_{21}&0\\
0&0\end{array}
\right)=\left(
\begin{array}{cc} -\partial_x\left(\wit{P}_{22}\right)^{-1}\partial_x&0\\
0&0\end{array}
\right)\,,
\end{equation}
with $\wit{P}_{22}=\partial_x \cdot g^\prime(\zeta)+g^\prime(\zeta)\cdot \partial_x$. This yields the reduced  Dirac Poisson tensor on the ``constrained" manifold of unidirectional right-moving waves,
\begin{equation}
\label{PDir-r}
P_{R}^D\equiv-\partial_x\left(\wit{P}_{22}\right)^{-1}\partial_x\, .
\end{equation}
Our final task is to compute, still in the MNL asymptotics, the inverse of the operator $\wit P_{22}$. 

A direct computation in this asymptotics shows that such an inverse is given  by the pseudo-differential operator
\begin{equation}
\label{p22-1}
({\wit P}_{22})^{-1}=\frac14\left( 2\partial_x^{-1}+\frac12\wit{B}\alpha\,(\partial_x^{-1}\,\zeta+\zeta\, \partial_x^{-1})-\frac18\wit{B}^2\alpha^2(2\partial_x^{-1}\zeta^2+2\zeta^2\partial_x^{-1}-\partial_x^{-1}\zeta\partial_x\zeta\partial_x^{-1}-\zeta\partial_x^{-1}\zeta)+\frac12\eps^2\wit{\kappa}\partial_x\right)\,,
\end{equation}
which yields, after some manipulation, the reduced Dirac  tensor
\begin{equation}
\label{PDi}
P_{R}^D=-\frac12\partial_x-\frac18 \alpha\wit{B}(\zeta \partial_x+\partial_x\, \zeta)+\frac1{32}{\alpha^2}\wit{B}^2(\zeta^2\, \partial_x+\partial_x\,\zeta^2+\zeta_x \partial_x^{-1}\zeta_x)-\frac14\eps^2\wit{\kappa}\partial_{xxx}\,.
\end{equation}
The Hamiltonian density reduces, on the constrained manifold $\sigma=\zeta-\frac14\,\alpha\,\wit{B}\,{\zeta}^{2}+\frac18\,{\alpha}^{2}{\wit{B}}^{
2}{\zeta}^{3}-\frac12\,{\epsilon}^{2}\wit{\kappa}\,\zeta_{{{\it xx}}}
$ 
and in the MNL asymptotics, to
\begin{equation}
\label{HD}
\wit{H}^D=\zeta^2+\frac14\alpha\wit{B}\zeta^3-\alpha^2\left(\frac3{32}\wit{B}^2+\frac12\wit{C}\right)\zeta^4\, . 
\end{equation}
Finally, it is easy to verify that the combination
\begin{equation}
P_{R}^D\dsl{\frac{\delta}{\delta\, \zeta} \int_\RR \wit{H}^D\, \D x}
\end{equation}
yields the unidirectional equation of motion (\ref{1direq}).

It is remarkable that, in the WNL asymptotics $\alpha=O(\eps^2)$, the operator $P_{R}^D$ yields the bi-Hamiltonian structure 
for the KdV equation. Indeed, 
\begin{equation}
\label{Magri}
P_{R,KdV}^D=\frac12\partial_x-\frac18 \alpha\wit{B}(\zeta \partial_x+\partial_x\, \zeta)-\frac12\,{\epsilon}^{2}\wit{\kappa}\,\partial_{{{\it xxx}}}
\end{equation}
can be written, after suitable rescaling of the variables, as 
\begin{equation}
\label{Magri2}
P_{R,KdV}^D=
\partial_x-\eps^2\left(\partial_{{{\it xxx}}}+ \zeta \partial_x+\partial_x\, \zeta\right),
\end{equation}
which is the Magri Poisson pencil for KdV, where the usual role of the pencil parameter is here played by the square of the inverse dispersion parameter  $\epsilon^{-2}$.

\section{Special solutions} 
\label{specsol}

In this 
section, we investigate some properties of the motion equations~(\ref{eqofmot}) which are relevant to their actual applicability as models of wave propagation, viz. their dispersive behaviour and 
their traveling wave solutions. \colr{This is a first step, which can be carried out without resorting to numerical methods,  necessary to assess the performance of the models we have derived with respect to established results for the parent Euler equations.}
\subsection{Linearization and the dispersion relation}
\label{displin}
Linearizing system~(\ref{eqofmot}) around the constant solution $(Z,S)$, $\zeta=Z+z(x,t)$ and  $\sigma=S+s(x,t)$ say, with the functions $z,s$ treated as infinitesimal, yields
\begin{equation}
\label{lineq}
\left\{\begin{split}
&z_t+As_x+ \alpha B (Z s_x+S z_x)-\alpha^2C (Z^2 s_x+ 2S Z z_x)+\epsilon^2 \kappa s_{xxx}=0
\\
&s_t+z_x+ \alpha B S s_x-\alpha^2C (S^2 z_x+2SZ s_x)=0
\end{split}\right.\, . 
\end{equation}
Looking for sinusoidal wave solutions of the form $(z,s)=(a_z,a_s)e^{i(kx-\omega t)}$ leads to the following algebraic eigenvalue  problem for the phase speed $c_p\equiv \omega/k$ as eigenvalue,
\begin{equation}
\label{matlin}
\left[ \begin {array}{cc} \alpha BS-2\alpha^2CSZ-c_p&\,\,\, A+\alpha BZ-\alpha^2CZ^2-\epsilon^2\kappa k^2
\\ \noalign{\medskip} 1-\alpha^2 CS^2  &\,\,\, \alpha BS-2\alpha^2CZS-c_p
\end {array} \right]
\left[ \begin {array}{c} a_z
\\ 
\noalign{\medskip} a_s\end {array} \right]
=\left[ \begin {array}{c} 0
\\ 
\noalign{\medskip} 0\end {array} \right]\,,
\end{equation}
giving the following dispersion relation
\begin{equation}
\label{disprel}
c_p= \alpha(BS-2\alpha CSZ) \pm \sqrt{(1-\alpha^2 CS^2)(A +\alpha BZ-\alpha^2 CZ^2-\epsilon^2\kappa k^2) }\, ,
 \end{equation}
whereby the critical threshold wavenumber  
\begin{equation}\label{criticone!}
k_c^2=(A +\alpha BZ-\alpha^2 CZ^2)/(\epsilon^2\kappa)
\end{equation} 
is identified, past  which the system becomes Hadamard ill-posed with $k>k_c$. Note that the factor $1-\alpha^2CS^2$ needs to be positive, for stability of long waves, i.e.,  $k\ll 1$, where the asymptotic model applies. This puts a bound on the admissible values of the equilibrium momentum shear $S$ at the critical values $S=\pm 1/(\alpha\sqrt{C})$, with $-1/(\alpha\sqrt{C})<S<1/(\alpha\sqrt{C})$. 
\colr{Such a bound, and in particular the fact that  the threshold wavenumber for~(\ref{exeom}) is independent of the magnitude of the shear $\sigma$, limits the applicability of this system in possible numerical applications. However,  it is inherent to our local Hamiltonian setting and will consistently recur in what follows, e.g. in the definitions of the domain of hyperbolicity of the dispersionless limit, as well as in the analysis of the travelling wave solutions of the dispersive case.}

With regard to the last point, it is worth noting that an asymptotic step, akin to the well know KdV to BBM near identity relation, can here be used advantageously \colr{to circumvent the hindrance to numerical applications due  the lack of well posedness of~(\ref{exeom}). Indeed, a} shift of the dependent variable $\sigma$, similar to~(\ref{sigbar}),
\begin{equation}
\bar{\sigma}\equiv \sigma+\epsilon^2 \bar{\kappa}\,\sigma_{xx} \Longrightarrow \sigma = \bsigma-\epsilon^2 \bar{\kappa}\,\bsigma_{xx}+O(\eps^4)
\,,
\label{bbmabcsh}
\end{equation}
where 
$
\bar{\kappa}={\kappa / A} \, , 
$
takes system~(\ref{exeom}) into the asymptotically equivalent form
\begin{equation}
\label{kaupsysreg}
\left\{\begin{split}
&\zeta_t+A \bsigma_x+ \alpha B(\zeta\bsigma)_x -\alpha^2 C (\zeta^2 \bsigma)_x=0
\\
&\bsigma_t+\zeta_x+ \alpha B \,\bsigma\bsigma_x -\alpha^2 C (\zeta \bsigma^2)_x=\epsilon^2\bar{\kappa}\bsigma_{xxt}
\end{split}\right.\, . 
\end{equation}
 
As hinted by the notation used, this step is equivalent to that of using layer averaged velocities, in defining the density weighted vorticity, instead of the velocities at the interface between layers used in the definition~(\ref{sigint}) of~$\sigma$. The dispersion relation for system~(\ref{kaupsysreg}) linearized around constant states $\bar\sigma=S$ and $\zeta=Z$ is readily obtained from modifying~(\ref{matlin}) above:
\begin{equation}
\label{matlinb}
\left[ \begin {array}{cc} \alpha BS-2\alpha^2CSZ-c_p&\,\,\, A+\alpha BZ-\alpha^2CZ^2
\\ \noalign{\medskip} 1-\alpha^2 CS^2  &\,\,\, \alpha BS-2\alpha^2CZS-c_p(1+\epsilon^2\bar{\kappa} k^2)
\end {array} \right]
\left[ \begin {array}{c} a_z
\\ 
\noalign{\medskip} a_s\end {array} \right]
=\left[ \begin {array}{c} 0
\\ 
\noalign{\medskip} 0\end {array} \right] \, .
\end{equation}
This is more resilient for well-posedness than~(\ref{disprel}). In fact, the eigenvalue $c_p$ is given 
by the solution of the quadratic equation
\begin{equation}
(1+\eps^2\bar{\kappa} k^2) c_p^2-q_1(2+\eps^2\bar{\kappa} k^2)c_p -q_2+q_1^2=0 \,,
\label{wlpb}
\end{equation}
where we have introduced the shorthand notation 
$$
q_1=\alpha BS-2\alpha^2CSZ\,, \quad q_2=(A+\alpha BZ-\alpha^2CZ^2)( 1-\alpha^2 CS^2)\,.
$$
\colr{The  asymptotic expansion of the discriminant of~(\ref{wlpb}) is  
$$
\Delta
=4\left(A+\alpha B Z - \alpha^2 C (A S^2+Z^2)+A \eps^2 \bar{\kappa}  k^2 \right)+O(\alpha^3,\alpha \eps^2)\,,
$$
which is certainly asymptotically  positive 
for values of $S$ and $Z$ of order~$O(1)$ with respect to the small~$\alpha$ parameter.  }


The standard dispersion relations for infinitesimal disturbances around  the quiescent state $\zeta=\sigma=0$ are obtained from the above relations setting $Z=S=0$, 
\begin{equation}
c_p^2= A -\epsilon^2\kappa k^2 \, ,
\end{equation}
and 
\begin{equation}
c_p^2= {A \over 1+\epsilon^2\bar{\kappa} k^2} \, ,
\end{equation} 
for systems~(\ref{exeom}) and~(\ref{kaupsysreg}) respectively. The role of the coefficient $A$ as the limiting long wave phase speed, and the different behaviors of these speeds in the large wavenumber $k$ limit, are especially transparent in this case. 

\remark{\rm
The analysis of the dispersionless counterpart of the system (\ref{eqofmot}) goes as follows (see \cite{CFO17} for the fully non-linear dispersionless case).
The dispersionless Hamiltonian density can be read off equation (\ref{ndEn}), 
\begin{equation}
H_d=\frac12\left(A\, {\sigma}^2+\alpha B\,  \zeta{\sigma}^2- \alpha^2 C\,  {\zeta}^2{\sigma}^2+{\zeta}^2\right)\,.
\end{equation}
Hence, the dispersionless equations can be written as
\begin{equation}
\label{eqd}
\begin{pmatrix}
\medskip \zeta_t\\
\sigma_t
\end{pmatrix}+
\begin{pmatrix}
\medskip H_{d,\, \zeta\sigma}&H_{d, \, \sigma\sigma}\\
H_{d,\, \zeta\zeta} &H_{d, \, \zeta\sigma}
\end{pmatrix}\, \begin{pmatrix}
\medskip \zeta_x\\
\sigma_x
\end{pmatrix}=\begin{pmatrix}
\medskip 0\\
0
\end{pmatrix}\, .
\end{equation}
The characteristic matrix of the system is, explicitly, 
\begin{equation}
\label{Dmat}
V=
\left(\begin {array}{cc} \alpha B\sigma-2\,\alpha^2 C\sigma\,\zeta&
-\alpha^2 C{\zeta}^{2}+\alpha B
\zeta+A\\ \noalign{\medskip}1-\alpha^2 C{\sigma}^{2}&\alpha B\sigma-2\,\alpha^2 C\sigma\,\zeta
\sigma\end {array} \right)\, ,
\end{equation}
and so the characteristic velocities are given by
\begin{equation}
\label{charvel}
v_{\pm}=\alpha B\sigma-2\,\alpha^2 C\sigma\,\zeta\pm\sqrt{(-\alpha^2 C{\zeta}^{2}+\alpha B
\zeta+A)( 1-\alpha^2 C{\sigma}^{2})} \, . 
\end{equation}
The hyperbolicity domain is thus the rectangular region in the hodograph space
\begin{equation}
(\zeta,\sigma)\in 
\left(
\frac { B-\sqrt {4\, C A+{B}^{2}}}{2\,\alpha C} ,
\frac { B+\sqrt {4\, C A+{B}^{2}}}{2\,\alpha C}
\right) 
\times 
\left(-\frac1{\alpha\sqrt{C}},\frac1{\alpha\sqrt{C}}\right)\,.
\end{equation}
}
\colr{{\rm
\remark{\rm
The regularization~(\ref{bbmabcsh}) is different from those used in~\cite{cbj,lanmin,BoMi14}, where the change of variables leading to the shear is done through the choice of a reference height of the horizontal velocities in the layers. The resulting models can be still ill-posed with respect to a dispersion critical wavenumber, however with an optimal choice of the reference height (typically, that of the bottom and top layer) the critical wavenumber can be made to maximize the well posed interval of the dispersion relation.}
}
}
{\rm
\subsection{Travelling wave solutions and their properties} 
Travelling waves for the $ABC$-system, 
\begin{equation}
\label{eqofmot3}
\left\{\begin{split}
&\zeta_t+\left(A {\sigma}+B{\zeta}{\sigma}- C{\zeta}^2{\sigma}+\kappa{\sigma}_{xx}\right)_x=0\\
&\sigma_t+\left({\zeta}+ \frac{B}{2} {{\sigma}^2}- C{\zeta}{\sigma}^2\right)_x=0\end{split}\right.\, , 
\end{equation}
rewritten here droppings stars and setting 
$\alpha=\eps=1$ in~(\ref{eqofmot}), 
are obtained via the ansatz $\zeta(t,x)=\zeta(x-c\,t), \sigma(t,x)=\sigma(x-c\, t)$ as the solution of the system
\begin{equation}
\label{TW1}
\left\{\begin{split}
&-c\zeta+A {\sigma}+B{\zeta}{\sigma}- C{\zeta}^2{\sigma}+\kappa{\sigma}_{xx}=K_1\\
&-c\sigma+{\zeta}+ {B \over 2}{{\sigma}^2}- C{\zeta}{\sigma}^2=K_2\end{split}\right.\, , 
\end{equation}
$K_1$ and $K_2$ being integration constants.
We limit ourselves to seek solitary wave solutions propagating into a quiescent state, i.e., $\zeta\to 0$ and $\sigma \to 0$ as $x\to 
\infty$, which sets $K_1=K_2=0$. The second equation in~(\ref{TW1})  yields the relation between $\zeta$ and $\sigma$,
\begin{equation}
\zeta=\sigma\, {c-B \sigma/2 \over 1-C \sigma^2}\,,
\label{zs}
\end{equation}
and substituting this into the first of~(\ref{TW1}) provides the quadrature formula
\begin{equation}
\kappa \sigma_x^2 =-\sigma^2\left(A-\frac14{(B \sigma-2 c)^2  \over1- C\sigma^2}\right)\,,
\label{sxeq}
\end{equation}
which can be interpreted as the mechanical analog of particle  of mass $2\kappa$ in a potential well $U(\sigma)$,
\begin{equation}\label{minus-U}
U(\sigma) \equiv \sigma^2\left(A-\frac14{(B \sigma-2 c)^2  \over 1-C\sigma^2}\right)\,.
\end{equation}
\begin{figure}[b]
\centering
{(a) \includegraphics[width=6cm]{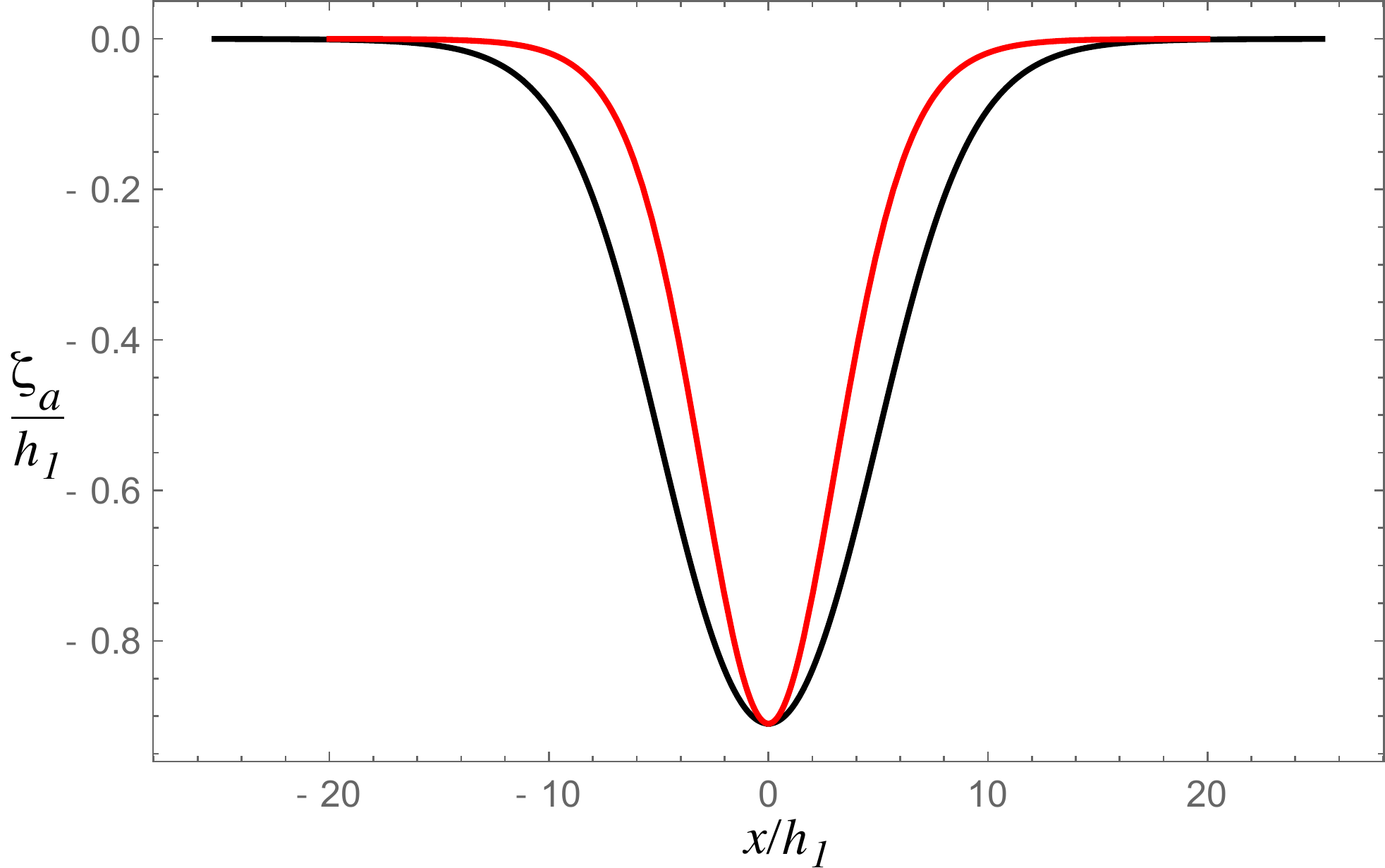}
\hspace{1cm}
(b)\includegraphics[width=6cm]{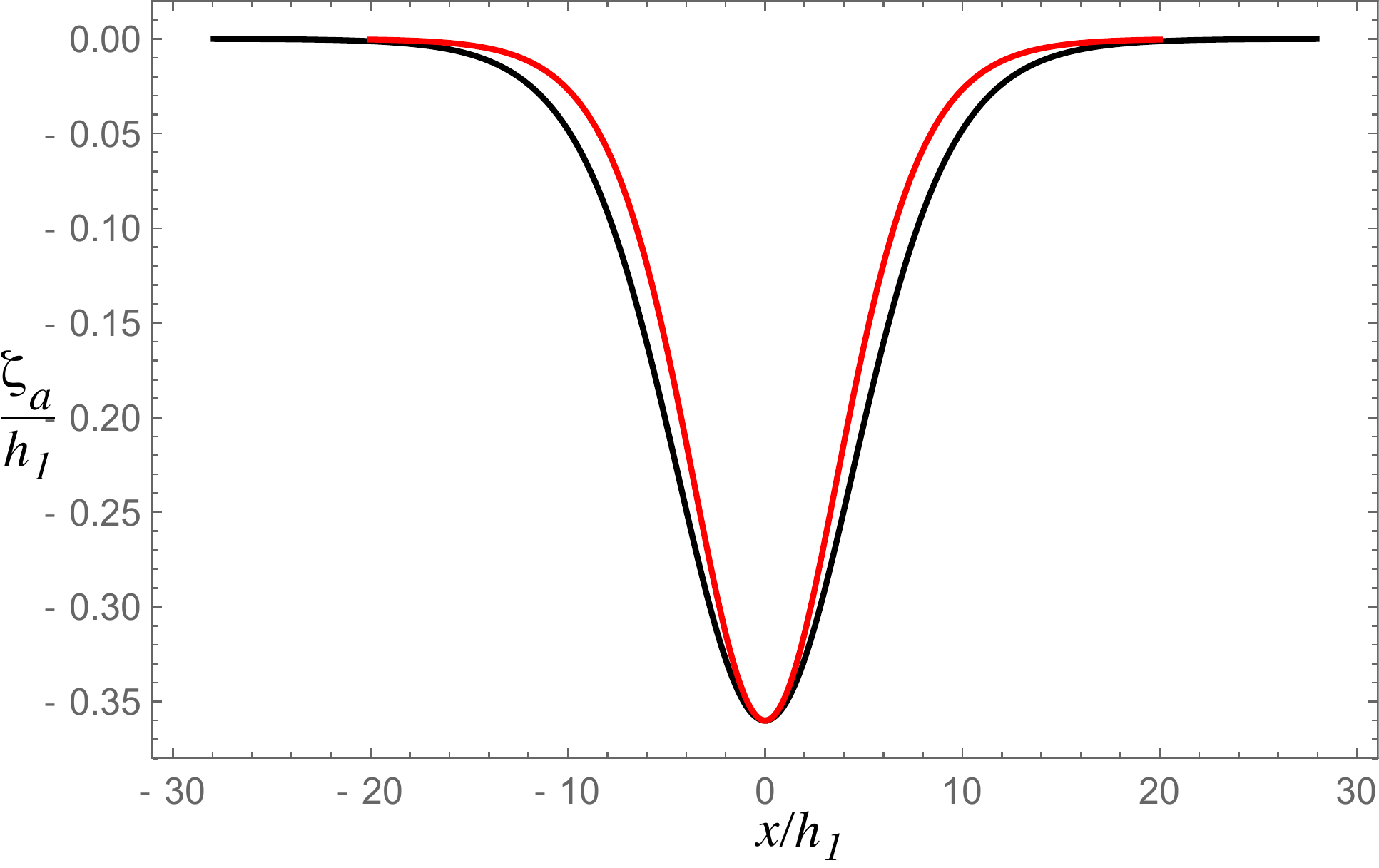}}
\caption{Comparison of internal solitary wave profiles $\zeta(x)$ of different amplitudes: red -- solution of~(\ref{eqofmot3}), black -- solution of strongly nonlinear model of~\cite{CC99}. Here the physical parameters are 
 $g=981$\,cm/s$^2$, $h_1=15$\,cm, $h_2=62$\,cm, $\rho_1=0.999$\,g/cm$^3$, $\rho_2=1.022$\,g/cm$^3$ (used in the experiment of~\cite{Grue}), and the maximum displacement is measured in units of the upper layer thickness $h_1$: (a) $\zeta_a=0.91 h_1$;  (b) $\zeta_a=0.36 h_1$.}
\label{compabcvsgnwav}
\end{figure}
An exact expression for $x$ as a function of $\sigma$ for the solution of~(\ref{sxeq}) can be found in terms of elliptic functions, but it is not particularly illuminating and will not be reported here. Once such an expression is obtained, its counterpart for the displacement $\zeta$ follows  immediately from~(\ref{zs}).
Typical wave solution profiles are shown in Figure~\ref{compabcvsgnwav}, where they are compared with those of the same amplitudes for the two-layer model of~\cite{CC99}; this in turn is known to provide good approximations to full Euler solutions in a broad range of physical parameters, including large nonlinearity. As this figure suggests, the differences between the two models become smaller for decreasing amplitudes, in agreement with the mild nonlinearity assumption underlying system~(\ref{eqofmot3}).   
\colr{
The potential (\ref{minus-U}) has been normalized to have a double zero for $\sigma=0$, and limits to $-\infty$ when $\sigma$ tends to $1/\sqrt{C}$ from the right and to $-1/\sqrt{C}$ from the left.  Solitary waves ---  always associated with  the null value of the energy of the corresponding mechanical system --- exist when  $\sigma=0$ is a local maximum for $U(\sigma)$, and $U(\sigma)$ has two more distinct non-zero roots $\sigma^*_{1,2}$ in the interval $(-1/\sqrt{C},1/\sqrt{C})$, so that $-U(\sigma)$ is
non-negative for $\sigma$ between $0$ and the smallest (in absolute value) of these additional roots (the limiting case of $\sigma^*_1\to \sigma^*_2$ corresponds to soliton solutions degenerating to  front-like solutions, as the orbit becomes heteroclinic). This implies that
(smooth)} solitary waves $(\zeta(x-c t),\sigma(x-c t))$ form a one-parameter family with respect to the speed $c$ in the interval 
\begin{equation}
 A<c^2<A+{B^2 \over 4 C} \, .
\end{equation}
Since $c_0\equiv \sqrt{A}$ can be interpreted  as the linearized speed of interfacial long-waves in a two-fluid system, this shows that nonlinear, solitary waves move faster than $c_0$ up to a limiting maximum speed defined by 
$$
c_m^2\equiv c_0^2+{B^2\over 4C}\,. 
$$  
The dependence on the speed parameter $c$ of the maximum displacements of solitary waves from equilibrium is (taking right-moving waves and $B>0$ to fix ideas)
\begin{equation}
\sigma_a={2 B c - 2 \sqrt{A\big(B^2-4C(c^2-A)\big)}\over B^2 + 4 A C} 
\,, \qquad
\zeta_a=
{-2A B +2c\sqrt{ A\big(B^2-4C(c^2-A)\big)}\over B^2-4 c^2 C}\,,
\label{zsac}
\end{equation}
respectively for $\sigma$  and  $\zeta$. (If $B<0$, and with right moving waves, the opposite sign of the square roots in the above formulae needs to be taken.) 
Regardless of the sign of $B$, 
at $c=c_m$ these waves  degenerate into fronts, with  amplitude of displacement  given by
\begin{equation}
 \sigma_m={B\over \sqrt{C(B^2+4AC)}} \,,  \qquad \zeta_m={B\over 2C}  \,. 
\end{equation}
Note that the sign of the $B$ coefficient in these relations determines whether the displacement $\zeta$ of internal solitary waves  is positive (waves of elevation) or negative (waves of depression), for $B>0$ and $B<0$, respectively. This sign is in turn determined by the value of the ratio $h_1/h_2$ with respect to the critical value~$\sqrt{\rho_1/\rho_2}$ (see Remark \ref{rmk:phys-par}).
Notice that for the WNL limit, which, as mentioned above, yields the Kaup-Boussinesq system~(\ref{kaupsys}) by formally taking $C=0$, the relation between wave amplitude and speed does not have a limiting extremum (at which the solitary waves degenerate into  fronts), and this relation is linear,
$$
\zeta_a=2 \,\sqrt{A}\,\,{c-\sqrt{A} \over B} \,. 
$$
\begin{figure}
\centering
{(a) \includegraphics[width=6.cm]{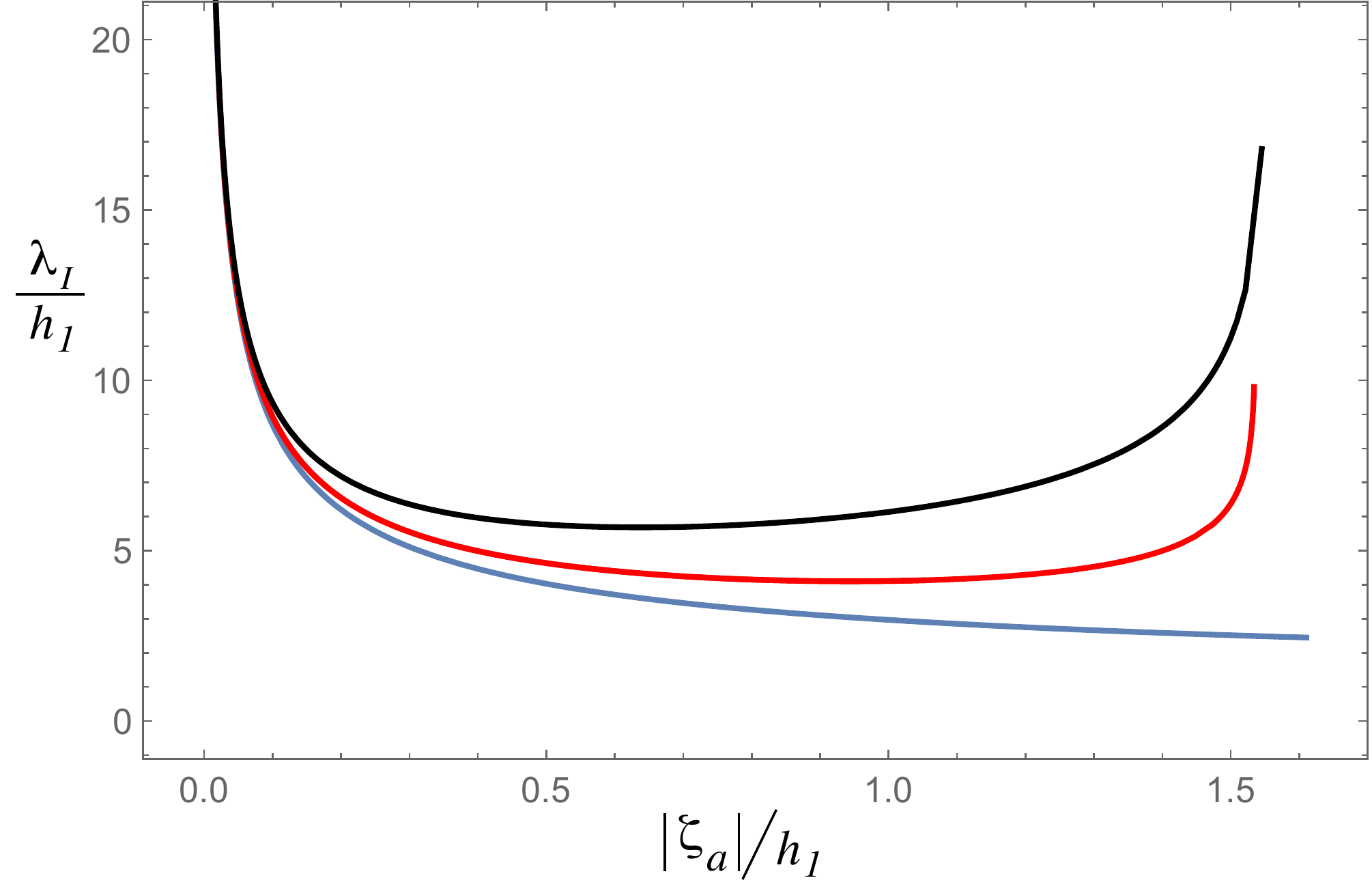}
\hspace{1cm}
(b)\includegraphics[width=6cm]{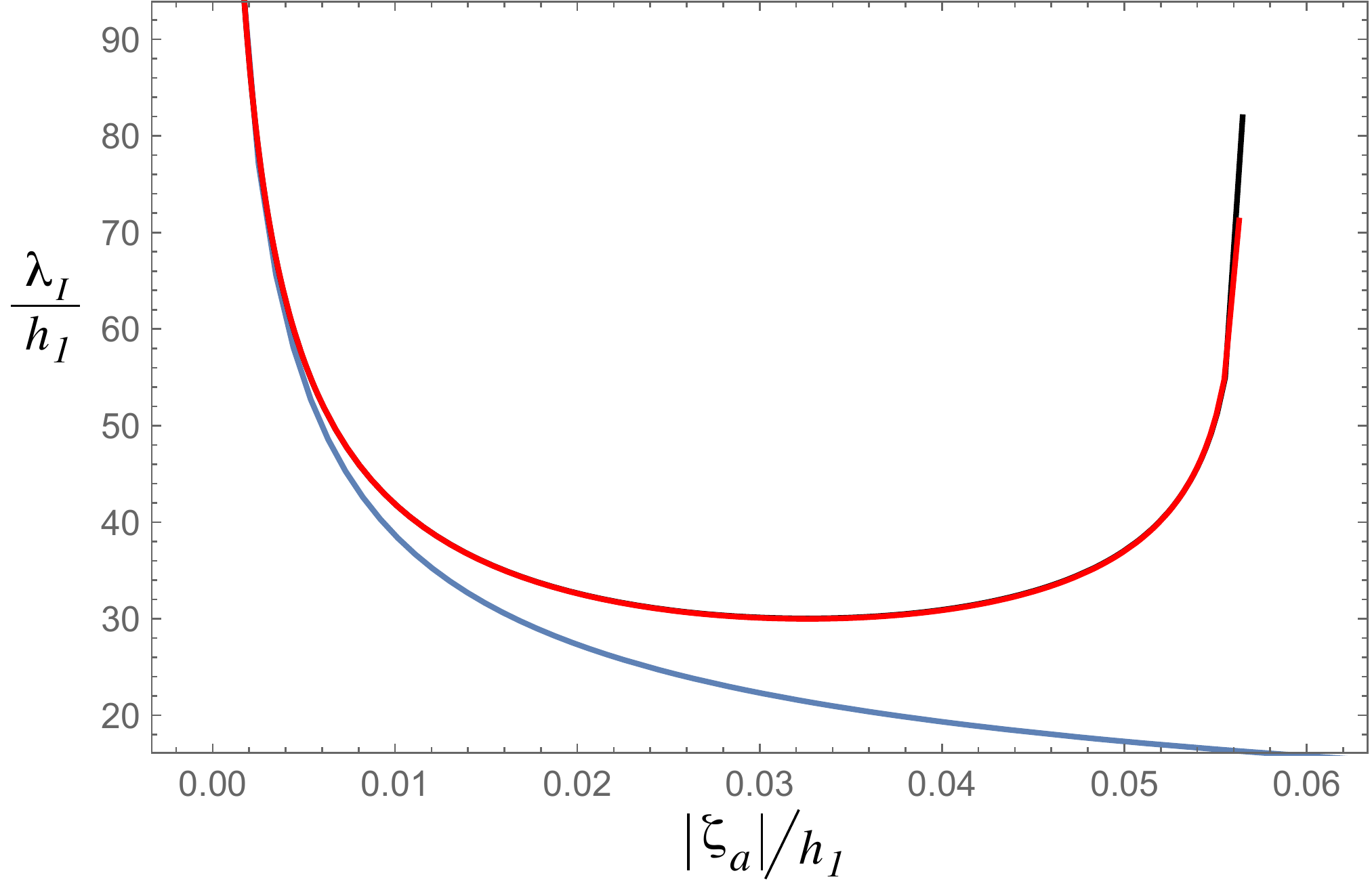}}
\caption{Effective wavelength vs.\ maximum displacement amplitude for internal solitary waves modeled by~(\ref{eqofmot3}) (red), by~(\ref{kaupsys}) (blue), and by the strongly nonlinear model of~\cite{CC99} (black). 
(a) $h_1=15$\,cm, 
$h_2=62$\,cm, 
$g=981$\,cm/s$^2$,
$\rho_1=0.999$\,g/cm$^3$, $\rho_2=1.022$\,g/cm$^3$.
(b) $h_1=55.1$\,cm, all other parameters as in (a). The axes are scaled with respect to the top layer thickness $h_1$. Note the different axis range in the two cases: for~(b) the amplitude range is much smaller and the waves are generally longer. Also note how  in both cases the location of the vertical asymptote  at the front value $\zeta_m$ is captured reasonably  well by model~(\ref{eqofmot3}), while the Kaup-Boussinesq system~(\ref{kaupsys}) lacks the front limiting case. As expected, for case~(a) which is far from the critical ratio, the intermediate amplitudes are off, while for the near-critical ratio case (b) the black curve is graphically indistinguishable from the red one.}
\label{lambdavsa}
\end{figure}

In terms of the ``hardware parameters" $h$'s and $\rho$'s the maximum speed $c_m$ and displacement amplitude $\zeta_m$ read, in the original dimensional variables, 
\begin{equation}
c_m^2=c_0^2\left(1+{(\rho_1 h_2^2  - \rho_2 h_1^2 )^2\over 4 (h_1+h_2)^2h_1 h_2 \rho_1 \rho_2}\right)\,,
\qquad \zeta_m={(\rho_1 h_2  + \rho_2 h_1 ) ( \rho_2h_1^2 - \rho_1 h_2^2)\over 2 \rho_1 \rho_2 (h_1+h_2)^2}\,,
\end{equation}
where, in dimensional form, the internal long-wave linear speed is given by 
$$
c_0^2 =g (\rho_2-\rho_1){h_1 h_2 \over \rho_1 h_2+\rho_2 h_1} \,. 
$$
As expected for the present asymptotic theory, carried out under the assumption of weak nonlinearity, these limiting values will in general be different from their exact counterparts of the two-layer Euler system \colr{which coincide  with the ones obtained with the fully non-linear model~\cite{CC99}}. The latter, $c_m^E$ and $\zeta_m^E$ say, are (see, e.g.,~\cite{CC99}) 
$$
(c_m^E)^2=c_0^2 \,{(h_1+h_2)(\rho_1 h_2 +\rho_2 h_1) \over h_1 h_2 (\rho_1+\rho_2+2\sqrt{\rho_1 \rho_2})} \,, 
\qquad 
\zeta_m^E={\rho_2 h_1^2-\rho_1 h_2^2 \over \rho_2 h_1+\rho_1 h_2+(h_1+h_2)\sqrt{\rho_1\rho_2}} \,,
$$
and they can be expected to be asymptotically close to $c_m$ and $\zeta_m$ as the critical ratio of depths and densities ${h_1\over h_2} = \sqrt{\rho_1\over \rho_2}$ is approached. \colr{This is due to the fact that the MNL model includes the  
$\alpha^2$-term which is dominant in this regime, since the coefficient of the term of order $\alpha$ vanishes in this limit}. Thus, the travelling solitary wave solutions of the present asymptotic model can be expected to provide a good approximation for their exact counterparts in the whole amplitude range, even approaching their front limit, when the depths and densities are such that the critical aspect ratio is approached to within an error of order~$O(\alpha)$; more precisely, it can be shown that
\begin{equation}
{h_1\over h_2} - \sqrt{\rho_1\over \rho_2}=\alpha \,\,\, \Longrightarrow \,\,\,
{c_m^E  - c_m \over c_m^E-c_0}=O(\alpha)\,, \quad \hbox{and} \quad {\zeta_m-\zeta_m^E \over \zeta_m^E}=O(\alpha) \,,
\label{critrh}
\end{equation}
in the limit $\alpha \to 0$, for which  $c_m^E,c_m \to c_0$ and $\zeta_m^E,\zeta_m\to 0$. 
These observations are exemplified by Figure~\ref{lambdavsa}  where the so-called effective wavelength
$$
\lambda_I\equiv {1\over \zeta_a}\int_0^{+\infty}\zeta(x)\, \D x 
$$
is plotted vs.\ $\zeta_a$ for two cases, one corresponding to the hardware parameters used in the experiment in~\cite{Grue}
and the other where the depth of the upper layer $h_1$ is adjusted to be close to the critical ratio as in~(\ref{critrh}) with $\alpha=0.1$. Figures~\ref{lambdavsa}, and its companion Figure~\ref{cvsa} where the so called nonlinear dispersion relation curve is also depicted, show a comparison with the analogous curves from the strongly nonlinear model~\cite{CC99}. 
 It is remarkable that the limiting values where solitary waves degenerate into fronts are somewhat accurately captured by the model even when these values fall well beyond the model's asymptotic validity. \colr{It is remarkable that this agreement occurs without recourse to an ad-hoc adjustment of the coefficients of the various term in the model, as done in~\cite{czb} in the context of a unidirectional reduction.}
 For the example provided in Figure~\ref{cvsa}(a), for instance, the limiting values are, respectively for the model and Euler systems, 
 $c_m/c_0=1.2565$ and $c_m^E/c_0=1.2580$,  and $\zeta_m/h_1=-1.5361$ and $\zeta_m^E/h_1=-1.5521$.
 Also, the nonlinear dispersion relation curve representing the wave velocity dependence on amplitude, $c=f(\zeta_a)$ with the function $f$ determined by the second equation in~(\ref{zsac}), remains close to that of the strongly nonlinear system  throughout the range of admissible displacement amplitudes, as seen in Figure~\ref{cvsa}(a). Once again, notice how all differences between models become graphically undetectable as the critical ratio is approached, as demonstrated by the (b) panels of these figures. 
\begin{figure}
\centering
{(a) \includegraphics[width=6cm]{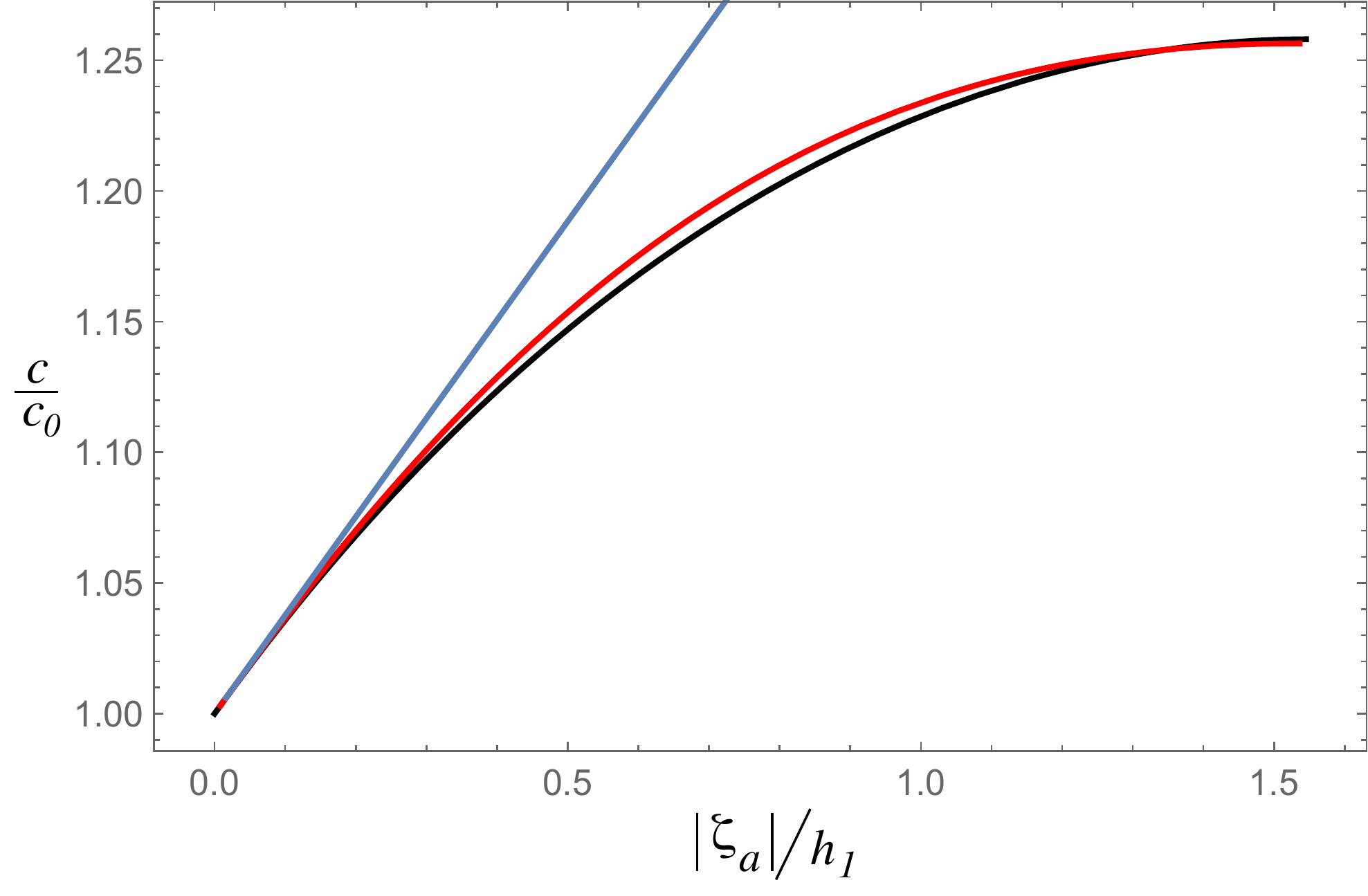}
\hspace{1cm}
(b)\includegraphics[width=6.6cm]{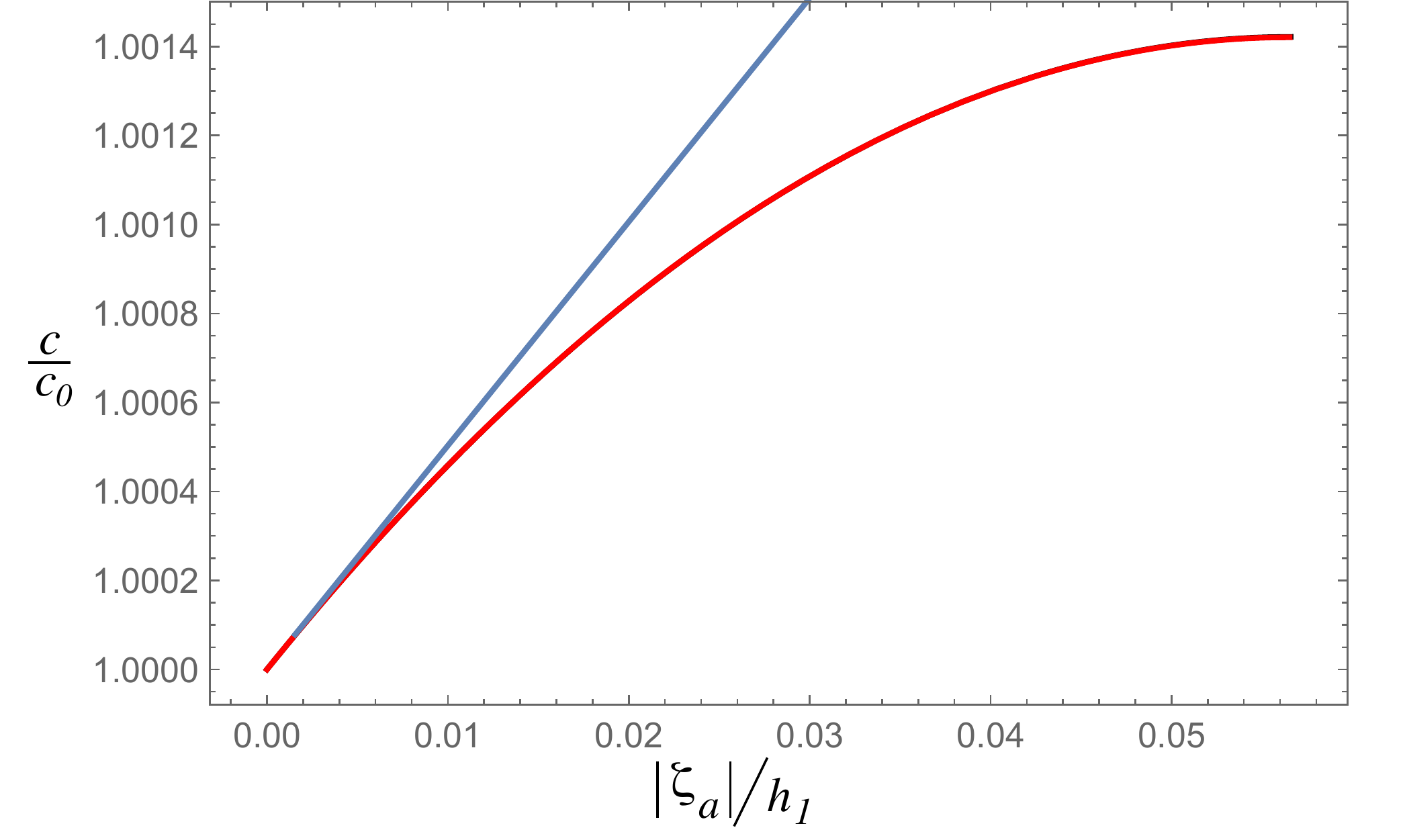}}
\caption{Same as Figure~\ref{lambdavsa}, but for speed of internal solitary waves vs.\ maximum displacement amplitude  modeled by~(\ref{eqofmot3}) (red), by~(\ref{kaupsys}) (blue), and by the strongly nonlinear model of~\cite{CC99} (black). Note that the discrepancy between~(\ref{eqofmot3}) and the strongly nonlinear model~\cite{CC99} is far less pronounced  for speed  than its effective wavelength counterpart, while the Kaup-Boussinesq~(\ref{kaupsys}) model is only asymptotically valid for small amplitudes in all cases. 
As for Figure~\ref{lambdavsa}, notice the different range of case (a) with respect to (b), and the undistinguishable overlapping between the red and the black curves.}
\label{cvsa}
\end{figure}

}

\section{Conclusions and perspectives}
\label{concper}

\rm 
We have \colr{applied} the technique of Hamiltonian reduction \colr{in its entirety, including the handling of constraints when present}, to derive model equations that inherit their structure from the parent Benjamin Hamiltonian formulation of a density stratified ideal fluid, under the asymptotic scalings of small amplitudes of fluid parcel displacements from their equilibrium  positions and under slow variations in their horizontal positions, i.e., long wave approximation and small dispersion.

The resulting main model generalizes to a bidirectional system the properties of the so-called Gardner unidirectional wave propagation equation, by allowing a quartic nonlinear term to enter the equations and \colr{ provide the necessary nonlinear contribution to} set a critical maximum displacement at which travelling solitary wave solutions degenerate into fronts, as well handle \colr{wave dynamics in a neighbourhood of critical depth ratio}. In contrast to its unidirectional counterpart, the additional nonlinearity within the bidirectional system makes wave properties such as the maximum amplitude $\zeta_m$ and the nonlinear dispersion relation $c(\zeta_a)$ \colr{close to their exact Euler counterparts, at least for the parameter range we have explored, even though the quartic term is formally subdominant to the other terms in the Hamiltonian with respect to the small asymptotic parameters carried by the coefficients. This is an unexpected feature of the MNL model that would have been difficult to anticipate based on the derivation alone. Of course, based on this metrics of travelling wave solutions the fidelity of the strongly nonlinear model~\cite{CC99} with respect to the parent Euler system remains unmatched. However, we should stress that the reasonable agreement is obtained here with a substantially simpler, local structure of the model. } Further, the Hamiltonian reduction techniques also allow for a systematic derivation of completely integrable models and in particular of Magri's \colr{bi-Hamiltonian structure}~\cite{Mag78} of the KdV equation from the parent Euler system, a program that fulfills the goal posed in~\cite{Olv84a,Olv84b} \colr{(for single layer fluids}).  

Future work will address reductions that are closer to the physical system by retaining higher order nonlinearity and dispersion, as well as remedy the drawbacks of ill-posedness injected by the asympotic truncations. The subtleties related to the double scaling limits with the two small parameters $\alpha$ and $\epsilon$ with respect to the physical hardware parameters densities $\rho$'s and depths $h$'s (see, e.g., \cite{BLS}), deserve further investigation which will be reported elsewhere.
\vspace{1cm}

\rm
\par\smallskip\noindent
{\bf Acknowledgments.}
We thank R. Barros, P. Lorenzoni, and R. Vitolo for useful discussions. \colr{Thanks are also due to the anonymous referees, for providing remarks and suggestions which helped improve the presentation, and for suggesting additional references}. 
This project has received funding from the European Union's Horizon 2020 research and innovation programme under the Marie Sk{\l}odowska-Curie grant no 778010 {\em IPaDEGAN}. We also gratefully acknowledge the auspices of the GNFM Section of INdAM, under which part of this work was carried out, and the financial support of the project MMNLP (Mathematical Methods in Non Linear Physics) of the
INFN. RC thanks the support by the National Science Foundation under grants RTG DMS-0943851, CMG ARC-1025523, DMS-1009750, DMS-1517879, DMS-1910824, and by the Office of Naval Research under grants N00014-18-1-2490 and DURIP N00014-12-1-0749. RC \& MP thank the Department of Mathematics and its Applications of the University of Milano-Bicocca for its hospitality. RC would also like to thank the Isaac Newton Institute for Mathematical Sciences, Cambridge, for support and hospitality during the programme HYD2 
where work on this paper was completed, with support by EPSRC grant EP/R014604/1.


\end{document}